\documentclass[12pt,letter]{article}
\usepackage{graphicx}
\usepackage{amsmath}
\usepackage{amssymb}
\usepackage{cancel}
\usepackage[left=2cm,right=2cm,top=3cm,bottom=2cm]{geometry}
\usepackage{setspace}
\usepackage{float}
\usepackage[super,sort&compress,comma]{natbib} 
\usepackage[font={normalsize}]{caption} 
\usepackage[labelfont=bf]{caption}
\usepackage{bm}
\usepackage[T1]{fontenc}
\usepackage{microtype}
\usepackage{subfigure}
\usepackage{chemformula}
\usepackage{tabularx}
\usepackage{booktabs}
\usepackage{multirow}
\usepackage{mathtools}
\usepackage{bm}
\usepackage{gensymb}
\usepackage{esvect}

\begin{document}
\title{\textbf{Learning Correlations between Internal Coordinates to improve 3D Cartesian Coordinates for Proteins}}
\date{}
\author{Jie Li$^{1}$, Oufan Zhang$^{1}$, Seokyoung Lee$^{1}$, Ashley Namini$^{2}$, Zi Hao Liu$^{2, 3}$, \\
João Miguel Correia Teixeira$^{2}$, Julie D Forman-Kay$^{2,3}$, Teresa Head-Gordon$^{1,4,5}$}
\maketitle
\noindent
$^1$Kenneth S. Pitzer Theory Center and Department of Chemistry, University of California, Berkeley, CA, USA\\
$^2$Molecular Medicine Program, Hospital for Sick Children, Toronto, Ontario M5S 1A8, Canada\\
$^3$Department of Biochemistry, University of Toronto, Toronto, Ontario M5G 1X8, Canada\\
$^4$Chemical Sciences Division, Lawrence Berkeley National Laboratory, Berkeley, CA, USA\\
$^5$Departments of Bioengineering and Chemical and Biomolecular Engineering, University of California, Berkeley, CA, USA\\
\begin{center}
corresponding author: thg@berkeley.edu
\end{center}

\begin{abstract}
\noindent
\textbf{Motivation}. We consider a generic representation problem of internal coordinates (bond lengths, valence angles, and dihedral angles) and their transformation to  3-dimensional Cartesian coordinates of a biomolecule. 
\vspace{2mm}


\noindent\textbf{Results}. We show that the internal-to-Cartesian process relies on correctly predicting chemically subtle correlations among the internal coordinates themselves, and learning these correlations increases the fidelity of the Cartesian representation. We developed a machine learning algorithm, Int2Cart, to predict bond lengths and bond angles from backbone torsion angles and residue types of a protein, and allows reconstruction of protein structures better than using fixed bond lengths and bond angles, or a static library method that relies on backbone torsion angles and residue types on a single residue.
\vspace{2mm}

\noindent\textbf{Availability and Implementation}. The Int2Cart algorithm has been implemented as a publicly accessible python package at https://github.com/THGLab/int2cart.

\end{abstract}

\newpage
\section*{\fontsize{12}{12}\selectfont INTRODUCTION}
\vspace{-3mm}
Biomolecular structures are described using two widely used mathematical representations: internal coordinates and Cartesian coordinates. The internal coordinate representation is defined by a set of bond lengths, bond angles, and dihedral or torsion angles, and provides a compact description in terms of the Z-matrix. In contrast, a Cartesian representation defines all of the atomic positions in Euclidean x,y,z coordinates and additionally captures the orientation of a molecule in space. Both representations are useful in certain contexts and applications. Internal coordinates can be beneficial for geometry optimizations\cite{Baker1999} and are the preferred description for NMR structure determination and refinement in which the bond lengths and bond angles are generally taken as fixed\cite{SCHWIETERS2001}, while Cartesian coordinates are the preferred format of molecular dynamics simulations\cite{Adcock2006} and X-ray crystallography structures provided in the Protein Data Bank (PDB) repository\cite{PDB2018}.

Figure 1 considers the internal coordinates of a protein backbone that contains the three torsion angles $\phi$ ($C-N-C_\alpha-C$), $\psi$ ($N-C_\alpha-C-N$), and $\omega$ ($C_\alpha-C-N-C_\alpha$), bond lengths $N-C_\alpha$ ($d_1$), $C_\alpha-C$ ($d_2$), and $C-N$ ($d_3$), and bond angles $N-C_\alpha-C$ ($\theta_1$), $C_\alpha-C-N$ ($\theta_2$), and $C-N-C_\alpha$ ($\theta_3$); side chain information that may effect the backbone could also include $C_\alpha-C_\beta$ ($r_1$) and $N-C_\alpha-C_\beta$ ($\alpha_1$) for example. When all of these quantities are specified exactly, the back-transformation from internal coordinates will also result in a perfect 3D Cartesian reconstruction of the protein backbone structure, using algorithms such as the natural extension reference frame (NeRF)\cite{nerf}. However, the Cartesian reconstruction is almost universally defined by only the backbone torsions while holding the bond lengths and angles fixed at mean values to decrease the complexity of structure modelling. Sometimes the variations on the $\omega$ torsion angles are also ignored and taken as fixed values of 0\degree or 180\degree, due to the planar nature of the peptide bond.\cite{craveur2013cis} One might assume that a protein structure can be reconstructed in Cartesian coordinates quite well utilizing fixed bond lengths and angles since they typically have quite small variations around their means. However, even small deviations from the mean of the stiff degrees of freedom can strongly influence the Cartesian reconstruction.   
\vspace{-3mm}
\begin{figure}[ht]
\begin{center}
\includegraphics[width=0.5\textwidth]{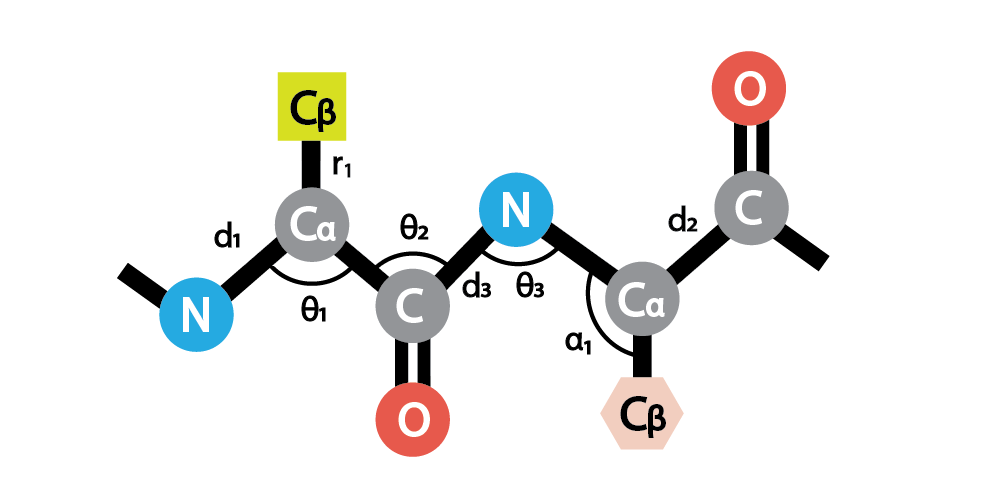}
\end{center}
\vspace{-10mm}
\caption{\textit{Schematic of the polypetide backbone and internal degrees of freedom}. Definition of the prediction targets: backbone bond angles $\theta_1-\theta_3$, backbone bond lengths $d_1-d_3$, $C_\alpha-C_\beta$ sidechain bond lengths $r_1$ and $N-C_\alpha-C_\beta$ sidechain bond angles $\alpha_1$.}
\label{fig:peptide}
\end{figure}
\vspace{-3mm}

\noindent
The origin of this error arises especially clearly from the nature of chain molecules: as the protein chain gets longer, small errors in bond lengths and bond angles can quickly accumulate and result in significant differences in the final back-transformed structure. According to a study by Holmes et al.\cite{holmes2004some} on globular proteins, the RMSD errors incurred in the internal coordinate back-transformations to C$_\alpha$ Cartesian positions under fixed bond lengths and angles is $\sim$6 \r{A} for an average 150-amino-acid protein, and can be as high as 40 \r{A} for larger proteins. 

Alternatively, one could replace the assumption of fixed bonds and bond angles with a statistical approach that conditions on sequence or structural correlations in the PDB. Given the many types of correlations that exist between the internal coordinates of globular proteins, such as the $\phi$ and $\psi$ torsion angles of the Ramanchandran plot\cite{RAMACHANDRAN196395} and the correlation between backbone and sidechain torsion angles used in the Dunbrack rotamer library\cite{dumbrack_library}, the correlations among the stiff bond lengths and angles with the flexible torsions should not be surprising. Earlier studies on the relationship between bond angles and $\phi$, $\psi$ torsion angles or amino acid types were mostly focused on the $N-C_\alpha-C$ bond angle, using both statistical methods and quantum mechanics calculation on model dipeptides.\cite{karplus1996experimentally, jiang1995predictions,schafer1995predictions,yu2001ab}. The work by Berkholz\cite{berkholz2009conformation} found that by using a static library for backbone bond angles dependent on backbone $\phi$, $\psi$ angles and residue types, the median RMSDs of protein reconstruction normalized to 100 amino acids is 2.85 \r{A}. Following this pioneering study, more recent work by Roberto and co-workers have extended the correlation to all bond lengths and bond angles centered on backbone N, C$_\alpha$ and C atoms.\cite{improta2015determinants,improta2015bond} Similarly, Lundgren et al. studied the correlation between protein backbone angles, secondary structure, and sidechain orientations.\cite{lundgren2012correlation}, and Ashraya et al. evaluated the steric-clash Ramanchandran maps conditioned on bond geometries. \cite{ravikumar2019stereochemical} However, none of these studies have considered the $\omega$ torsion angle dependence of the internal coordinates.

This work provides a more comprehensive machine learning approach that both quantifies and learns internal coordinate correlations within a deeper amino acid sequence context, that in turn provides a more accurate prediction of the 3D Cartesian coordinates relative to the errors incurred under the standard assumptions of fixed bond lengths and angles. By capturing the subtle correlations observed among internal coordinates, the Int2Cart (Internal to Cartesian) algorithm reduces the reconstruction RMSD error to $\sim$2.3 \r{A} for test proteins less than 150-amino-acids, and an average RMSD of $\sim$3.5 \r{A} over the entire test set for globular proteins as large as 543 amino acids. In addition we consider the "out-of-distribution" performance of Int2Cart on two more difficult test systems, including 42 proteins obtained from the 12th Community Assessment of Structure Prediction (CASP-12) as well as accurate reconstructions for intrinsically disordered protein (IDP) ensembles that we find generates lowers errors and fewer undesirable steric clashes. The Int2Cart deep learning model developed for proteins should be general to other types of chain molecules including RNA, DNA, and lipid molecules if appropriate data sets are made available. 

\section*{\fontsize{12}{12}\selectfont METHODS}
\vspace{-3mm}
\textit{Dataset preparation.} We have used SidechainNet\cite{King2021} as a preprocessed dataset that uses clustering techniques to extract protein sequences and structures with defined similarity cutoffs, to reduce bias in the original PDB structures, and to prevent information leakage from the training set to the test set relevant to assessing the machine learning generalization.\cite{AlQuraishi2019,King2021} The SidechainNet dataset represents each protein by its amino acid sequence, backbone and sidechain torsion angles ($\phi$, $\psi$, $\omega$, $\chi_1$, $\chi_2$, etc), backbone bond angles $\theta_1$-$\theta_3$, as well as the all-atom 3D coordinates. For this study we ignore the sidechain torsions as we are only reconstructing backbones, and also supplement the protein dataset with backbone bond lengths $d_1$-$d_3$ calculated from the 3D coordinates for training, validation and test sets. We also identified some $\theta_{2}$ and $\theta_3$ bond angles that were incorrect due to missing atoms in the next residue, and they were masked out along with the residue at the end of the protein chain. We combined validation sets from 10\% to 50\% similarity cutoffs to serve as our primary test data.\cite{AlQuraishi2019}  When needed, we separated test set proteins at the broken chain positions and only retained chains longer than 50 consecutive amino acids.\cite{AlQuraishi2019} Additionally we compiled two alternative and more difficult test data sets to validate the transferability of our model: the prediction targets from CASP-12 as well as reconstructions for IDP structural ensembles.

Our final training dataset contained 41,380 proteins with a minimum sequence length of 20 and maximum length of 4914 amino acids. The primary test set was comprised of 140 protein or protein fragments with sequence lengths between 23 and 543 amino acids. The challenging 42 proteins or protein fragments in the CASP12 test set ranged in length from 51 to 599 amino acids. The IDP ensemble comprised 1000 conformations for the N-terminal 92 residues of the Sic1 protein;\cite{Mittag2008,Gomes2020} we extracted the backbone torsions and rebuilt the Cartesian structures for each conformer under different assumptions about the bond lengths and bond angles as reported in Results.

\begin{figure}[ht]
\begin{center}
\includegraphics[width=0.55\textwidth]{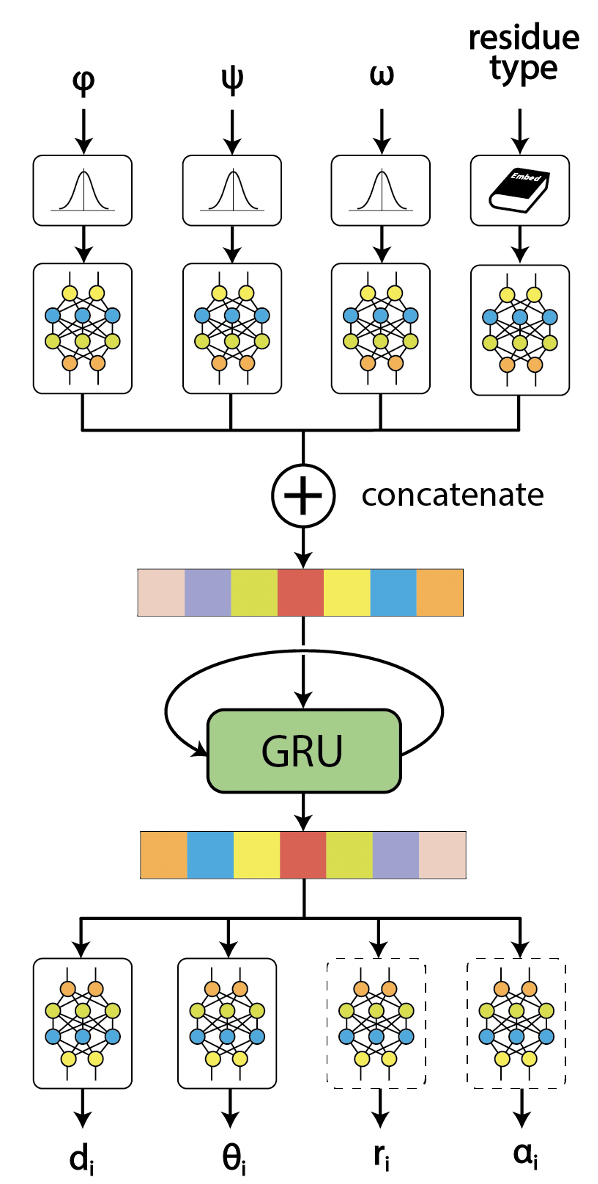}
\end{center}
\vspace{-7mm}
\caption{\textit{Schematic of the Int2Cart neural network architecture}. The neural network is a gated recurrent unit (GRU) recurrent neural network. The inputs at each timestep are the concatenated latent vectors from Gaussian-smeared $\phi$, $\psi$ and $\omega$ torsion angles and embedded residue types; variations on the Int2Cart network can include the use of $\chi$ sidechain angles as well. The latent vector output from GRU are connected with multiple output networks to predict different targets.}
\label{fig:GRU}
\end{figure}

\vspace{2mm}
\noindent
\textit{Neural network design.} The structure of the deep neural network, Int2Cart, is depicted in Figure \ref{fig:GRU}. We utilized 3 layers of stacked bidirectional gated recurrent units (GRU) as the central component, each of which contains a hidden state $h_t$ with its information updated by the reset and update mechanisms for each element in the input sequence through the following set of equations\cite{gru}:
\vspace{-2mm}
\begin{align}
    r_t&=\sigma(W^rx_t+U^rh_{t-1}+b^r) \\
    z_t&=\sigma(W^zx_t+U^zh_{t-1}+b^z)  \\
    \Tilde{h}_t&=\tanh{(W^nx_t+b^{nx}+r_t\odot(U^nh_{t-1}+b^{nh}))} \\
    h_t&=(1-z_t)\odot\Tilde{h}_t+z_t\odot h_{t-1}
\end{align}
where $[W^r,W^z,W^n,U^r,U^z,U^n,b^r,b^z,b^{nx},b^{nh}]$ are the trainable parameters of the model, $x_t$ is the input to the cell at the current timestep, and $r_t$ and $z_t$ represent the reset and update gates, which are numbers between $(0,1)$ that control how much information to retain in the new update vector $\Tilde{h}_t$ and how the new hidden state vector $h_t$ is composed from the update vector $\Tilde{h}_t$ and the old hidden state $h_{t-1}$. $\sigma$ denotes the sigmoid function, and $\odot$ represents element-wise multiplication. Dropout was applied to the hidden states between layers, so that $x_t^{(l)}=h_t^{(l-1)}\odot\delta_t^{(l-1)}$, where each $\delta_t^{(l-1)}$ is a Bernoulli random variable that zeros out elements in the hidden state vector with a probability defined by the dropout rate.  

The inputs into the first layer of GRU cells are the $\phi$, $\psi$ and $\omega$ torsion angles and the amino acid types. Each torsion angle, $a$, was represented by a Gaussian smearing function discretized to a vector of length 180 to account for uncertainty in the data, denoted $x_{ia}$
\begin{equation}
    x_{ia}=\exp\left(-\frac{\rm diff(\alpha_i,\hat{x}_a)}{2\sigma^2}\right) \\
\end{equation}

\noindent
where $\alpha_i$ is the actual $\phi$, $\psi$ or $\omega$ angle and $\rm \hat{x}_a=(-180+2*a)$ (both in degrees), and in this work we used $\sigma=0.5\degree$. The custom $\rm diff$ function 
\begin{equation}
    \rm diff(\alpha_i,\hat{x}_a)=\min(|\alpha_i-\hat{x}_a|, \min(|\alpha_i-\hat{x}_a-360\degree|,|\alpha_i-\hat{x}_a+360\degree|))
\end{equation}
ensures that the periodicity of the angles is taken into account.  Each smeared torsion angle vector is further transformed through two fully-connected layers with 90 and 64 units each and Rectified Linear Units (ReLU) activation\cite{relu} to generate latent representations of the torsion angles. The residue types are encoded by a trainable embedding dictionary and formulated into latent vectors of length 64.\cite{} The torsion angle latent vectors and the embedded residue types are then concatenated and transformed together through 2 fully-connected layers with 128 and 64 units and ReLU activation and constitute the inputs into the GRU cells.

The hidden state output from the last GRU layer is connected with multiple outputs to predict the backbone bond lengths and bond angles, or optionally sidechain bond lengths and bond angles as well. Each output is a fully-connected neural network with a hidden layer of size 100 and activation ReLU, and the output has size of 1 without any activation. The raw outputs are scaled by the standard deviation and translated by the mean value of that data type in the training dataset. The means and standard deviations we used are provided in Supplementary Table 1.
\noindent

\vspace{3mm}
\noindent
\textit{Training details of the Int2Cart machine learning method.} The neural network was trained by minimizing the weighted mean square error loss function
\begin{equation}
    \mathcal{L}=\sum_{i}w_i(y_i-\hat{y_i})^2
    \vspace{-2mm}
\end{equation}

\noindent
where $w_i$ controls the weighting for different data types in the loss function, $y_i$ are the predictions from the model and $\hat{y_i}$ are the actual values from the data set. In practice we used the same weighting for all the data types. Missing data targets were masked out during the training. We used the Adam optimizer\cite{Kingma2015} with an initial learning rate of 0.001 and an exponentially decayed learning rate schedule, so $ \mathrm{lr}_i=\exp(-i*\alpha)$ where $i$ is epoch number and $\alpha=0.05$ in our case. The model was trained for a total of 100 epochs using a batch size of 128.

\vspace{4pt}
\noindent
\textit{Building all-atom Cartesian structures from internal angle model predictions.} With the full profile of backbone torsion angles and predictions of bond lengths and bond angles from the model, the 3D Cartesian structure of the protein containing all backbone atoms is reconstructed using the SidechainNet package.\cite{King2021} It utilizes the natural extension reference frame (NeRF) algorithm\cite{nerf} to sequentially calculate the position of the next atom with the positions of three previous atoms and the new bond length, bond angle and torsion angle. The all-atom backbone Cartesian structures for all the protein fragments in our test data set are built from either the Int2Cart algorithm vs. a standard baseline of using fixed bond lengths and bond angles (Fixed) with no conditioning on torsions or amino acid type. 

\section*{\fontsize{12}{12}\selectfont RESULTS}
\begin{figure}[ht]
\begin{center}
\includegraphics[width=0.95\textwidth]{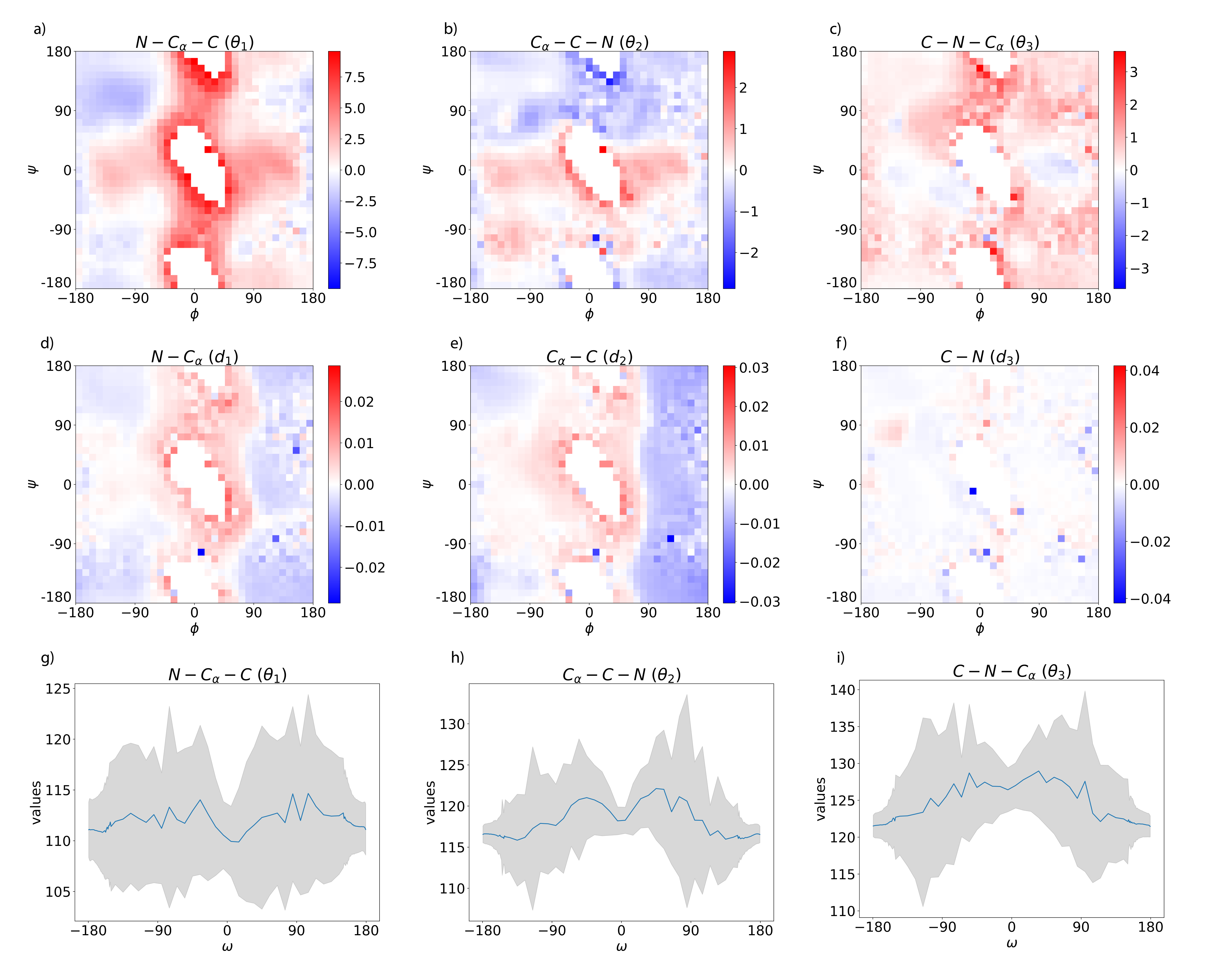}
\end{center}
\vspace{-10mm}
\caption{\textit{Variations of bond angles and bond lengths as a function of ($\phi$, $\psi$), or $\omega$ torsion angles}. a-f) Bond angle and bond length deviations from the mean values averaged over $\phi$ and $\psi$ angles of the training set. The regions of red correspond to wider angles and longer bonds while the region in blue show reduced angle and bond values relative to the mean. The bond lengths and bond angles were categorized according to $\phi$ and $\psi$ angles rounded to the closest tens, and the data are aggregated by calculating the means and standard deviations in each bin. The standard deviations are provided in Figure S1. g-i) Mean values and standard deviations of bond angles as a function of $\omega$. The blue solid line represents mean values of bond angles at specific $\omega$ torsion angles, and the gray regions correspond to one standard deviation.
}
\label{fig:dist}
\end{figure}

\vspace{-3mm}
\textit{Statistical analysis of the protein training set.} Given the large collection of protein structures deposited into the PDB, we first consider a statistical analysis of protein bond lengths and bond angles when analyzed over the training set. Overall the distributions of these internal coordinate values are mostly Gaussian with relatively small standard deviations of $\sim$ 0.01 \r{A} for bond lengths and $\sim$ 2.6$\degree$ for bond angles. Figure \ref{fig:dist} a-f depicts the deviations from the mean bond length and angle values for a given $(\phi,\psi)$ combination, and confirms the existence of strong correlations among the internal coordinates averaged over the training data set. Specifically, the $\theta_1$ angle is larger for the right-hand and left-hand helix regions in the Ramanchandran plot, while the beta-sheet regions have more narrow $\theta_1$ angles, with deviations from the mean as large as 7.5\degree. The $\theta_2$ values are strongly correlated with the $\psi$ torsion angle, with larger angles when $\psi$ is between $-100\degree$ and 0\degree, and smaller angles than the mean otherwise. The $\theta_3$ values for nearly all of the $(\phi,\psi)$ combinations are larger than average, but have smaller angles for helix regions. Similarly, the $d_1$ and $d_2$ bond lengths show greater correlations with the $\phi$ torsion angle, with a preference for larger values when $\phi$ is between $-50\degree$ and $+50\degree$, in which the bond lengths change by as much as 0.02 \AA. Finally the correlation for the peptide $d_3$ bond with the backbone torsions is weak, consistent with its partial double bond character, except for a few hot spots where it can vary up to 0.04 \AA. These correlations are statistically meaningful, because the standard deviations in each bin are smaller than the mean value differences (Supplementary Figure 1), which means the statistical bias is more significant than the variance.

We have also considered the relationships between backbone $\omega$ torsion angles with bond angles (Figure \ref{fig:dist} g-i) and bond lengths (Supplementary Figure 2), and found interesting correlations between internal coordinates and $\omega$ torsion angles. The majority of peptide bonds in proteins are in \textit{trans-} conformation, with $\omega$ torsion angles close to 180\degree. However, \textit{cis-} peptide bonds tend to associate with smaller $\theta_1$ angles and larger $\theta_2$ and $\theta_3$ angles. This result also makes structural sense since \textit{cis-} peptide bonds incur more steric repulsion between sidechains of two consecutive residues, and larger $\theta_2$, $\theta_3$ and smaller $\theta_1$ values allow the sidechains to be more separated. On the other hand, the correlation between bond lengths and $\omega$ torsion angles are not obvious (Supplementary Figure 2). These correlations dependent on $\omega$ are also important for accurately predicting internal coordinates from backbone torsion angles, as we will show later.

\begin{figure}[ht]
\begin{center}
\includegraphics[width=0.95\textwidth]{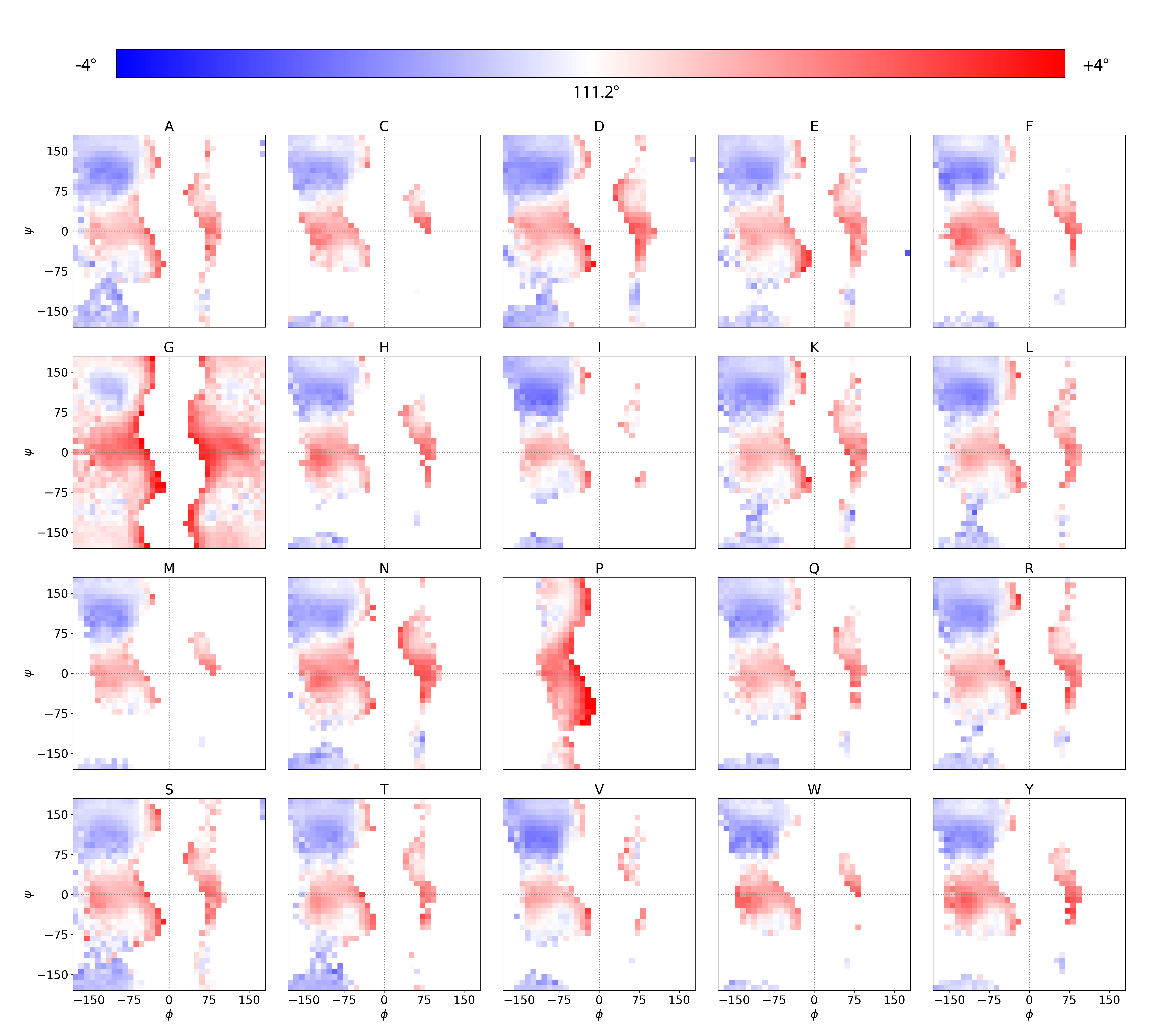}
\end{center}
\vspace{-7mm}
\caption{\textit{$N-C_{\alpha}-C$ bond angle deviations from the mean values averaged over $\phi$ and $\psi$ angles as a function of residue type}. The regions of red correspond to longer bonds while the region in blue show reduced bond values relative to the mean. The $N-C_{\alpha}-C$ bond angles were categorized according to $\phi$ and $\psi$ angles rounded to the closest tens.}
\label{fig:theta1-res}
\end{figure}

When we consider the observed distributions of all six internal coordinates conditioned on the residue type (Supplementary Figure 3), we find that the distributions are quite similar between amino acids with only subtle differences in the shape of the peaks, with the exception of glycine, which tends to have $d_1$ and $d_2$ values that are smaller, and $\theta_1$ angles that are larger than other residues. Proline also defines an exception, with larger $d_1$ values due to the formation of the five-membered ring that requires longer bond lengths. However the bond length and angle distributions when conditioned on backbone torsions and residue type exhibit notable variations across \textit{all} twenty amino acids as seen in Figure \ref{fig:theta1-res} for the $\theta_1$ bond angle, as well as for the other backbone bonds and angles provided in Supplementary Figures 4-8.

\textit{Machine learning of sequence and structural correlations}. While the correlation graphs just described could serve as a source for bond lengths and angles when backbone torsion angles and residue types are provided, we are still missing the deeper sequence correlations that are buried beneath the statistics of the single residue results. Therefore, we trained a deep neural network on the same data in order to capture the more subtle correlations among the internal coordinates conditioned on amino acid sequence. After training, the Int2Cart neural network was used to predict the primary test set which has at most 50\% sequence and structural similarity with the training proteins. The results are summarized in Table 1 in terms of root-mean-square error (RMSE) and Pearson correlation coefficients ($R$). We find that the bond length predictions are within computational uncertainty as measured by the variance in the data set, while predictions on the bond angles are more successful in terms of the RMSEs that are smaller than the dataset variance.  

\begin{table}[ht]
    \centering
\caption{\textit{Int2Cart prediction accuracy on backbone bond lengths and bond angles}. Accuracy is assessed in terms of root-mean-square error (RMSE) and Pearson correlation coefficients ($R$).}
    \begin{tabular}{lcc}
        \hline\hline
            \textbf{Data type}& RMSE & $R$     \\ \hline
        
        \textbf{N-C$_\alpha$-C  ($\theta_1$)} & 1.90\degree & 0.71 \\
        \textbf{C$_\alpha$-C-N ($\theta_2$)} & 1.08\degree & 0.44\\
        \textbf{C-N-C$_\alpha$  ($\theta_3$)} & 1.36\degree & 0.54\\
                           
        \hline
        \textbf{N-C$_\alpha$ ($d_1$)} & 0.011 \AA  & 0.38 \\ 
        \textbf{C$_\alpha$-C ($d_2$)} & 0.011 \AA  & 0.36  \\
        \textbf{C-N ($d_3$)} & 0.007 \AA & 0.50 \\  \hline
    \end{tabular}

\label{tab:prediction_accuracy}
\end{table}

\begin{figure}[ht]
\begin{center}
\includegraphics[width=0.98\textwidth]{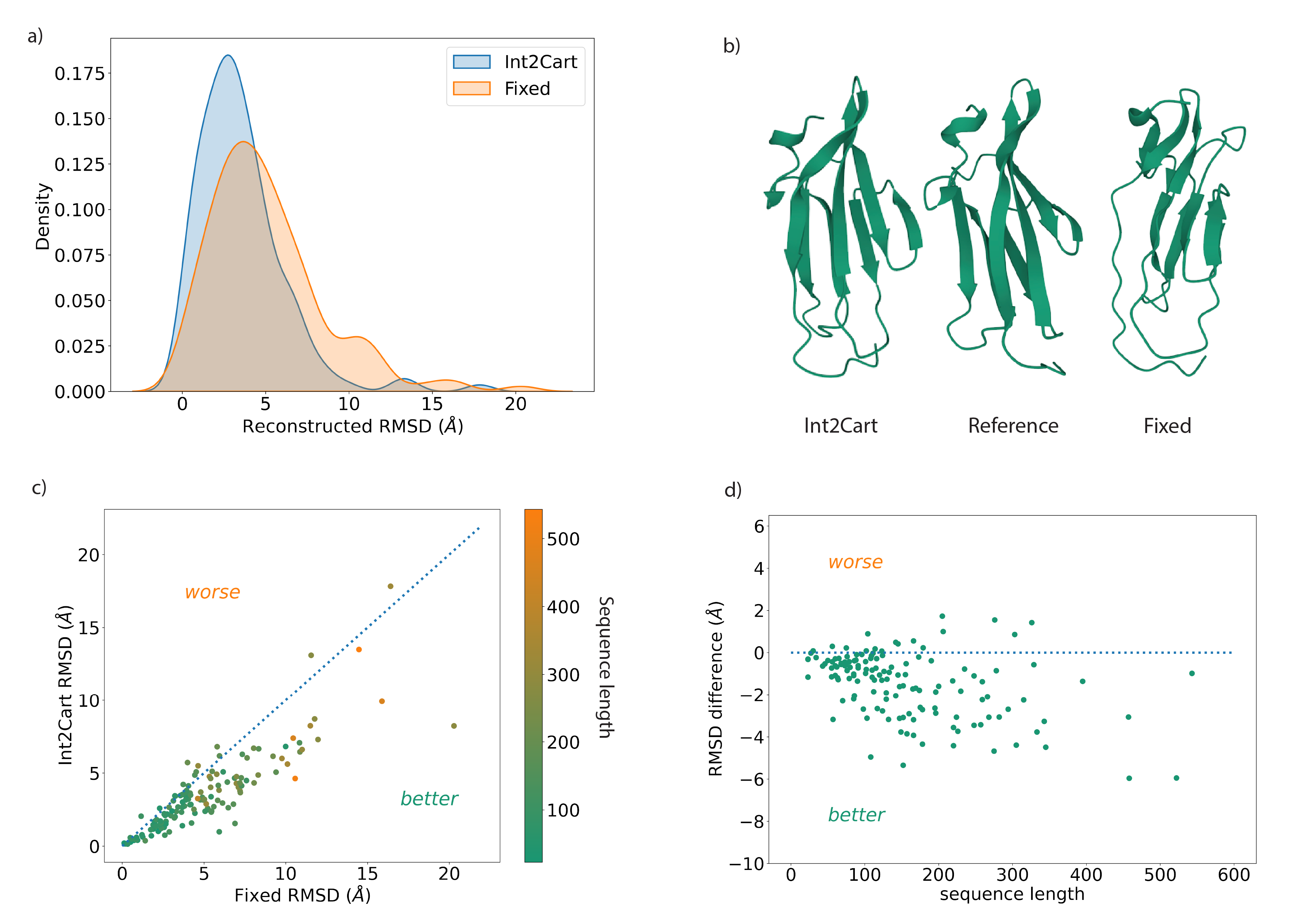}
\end{center}
\caption{\textit{Comparison of 3D Cartesian reconstructions of test set proteins using Int2Cart predictions and compared to fixed bond distances and angles.} (a) Distribution of the RMSD in reconstructed Cartesian coordinates using bond length and bond angle models learned from Int2Cart compared to Fixed values. (b) An example of the backbone representation using the different bond length and bond angle models. (Protein structure: TBM0872 from CASP12, PDBID 5JMB\cite{5jmb_structure})(c) Comparison of Cartesian reconstruction error between Int2Cart and Fixed relative to the reference structure. (d) Improvement of Int2Cart over Fixed as a function of amino acid length.}
\label{fig:results-rmsd}
\end{figure}

\textit{Cartesian coordinate reconstructions}. Given the three torsion angles $[\phi,\psi,\omega]$ for each residue over the entire protein sequence as input, we next consider how well the Cartesian coordinates are reconstructed based on whether bond and angle geometries are held fixed or learned from Int2Cart. Figure \ref{fig:results-rmsd}a reports the distribution of root-mean-square deviations (RMSD) for all backbone atoms with respect to the actual PDB structure for all proteins in the test set using Int2Cart and the Fixed method. In general, the reconstructed RMSD distributions for the Int2Cart structures are centered around lower RMSD values compared with the Fixed structures, with a median RMSD of 2.14 \r{A} when normalized to 100 amino acids, and the average RMSD over the entire test set is 3.5 \r{A}. By contrast the Fixed model yields a median RMSD of 3.24 \r{A} when all proteins are normalized to 100 amino acids, and the average over the entire test set is 5.1 \r{A}. The Int2Cart results are notably better than previous results, including the Berkholtz study\cite{berkholz2009conformation} that conditioned on $\phi$, $\psi$ and amino acid type, showing that deeper sequence correlations are beneficial to learn.

Figure \ref{fig:results-rmsd}b illustrates that the proteins reconstructed by assuming fixed bond lengths and bond angles have lost significant secondary structure integrity compared to the reference structures, whereas the Int2Cart structures retained a much higher proportion of secondary structural elements. The pairwise RMSDs for the test proteins reconstructed by Int2Cart are compared to those reconstructed using the Fixed approach in Figure \ref{fig:results-rmsd}c, and the RMSD difference between these two methods as a function of sequence length is shown in Figure \ref{fig:results-rmsd}d. It is evident that the vast majority of the test set proteins benefit from the machine learned bond lengths and bond angles, with an average improvement of $2-4$ \r{A} RMSD over using Fixed bond lengths and bond angles. There is no obvious correlation between the RMSD improvements made by Int2Cart over Fixed in regards sequence length, although the largest improvements occur in those proteins with longer amino acid sequences.  

Beyond the anecdotal case in Figure \ref{fig:results-rmsd}b, we performed more extended analysis of Int2Cart and Fixed performance in regards the radius of gyration (Rg) and secondary structure recovery rate (SS-match) over the whole test dataset. Although we find that the Int2Cart Cartesian predictions have closer R$_g$ values to the ground truth structures, the Fixed Cartesian structures still yields a comparably good result as seen in Supplementary Figure 9. However Figure \ref{fig:newfig}a shows that Int2Cart systematically improves upon Fixed in regards the SS-match values with a higher proportion of test set proteins that have SS-match values larger than 0.8 for Int2Cart predicted structures (Figure \ref{fig:newfig}b), which translates to more than 80\% of the residues having correct secondary structure assignments. 

\begin{figure}[ht]
\begin{center}
\includegraphics[width=0.98\textwidth]{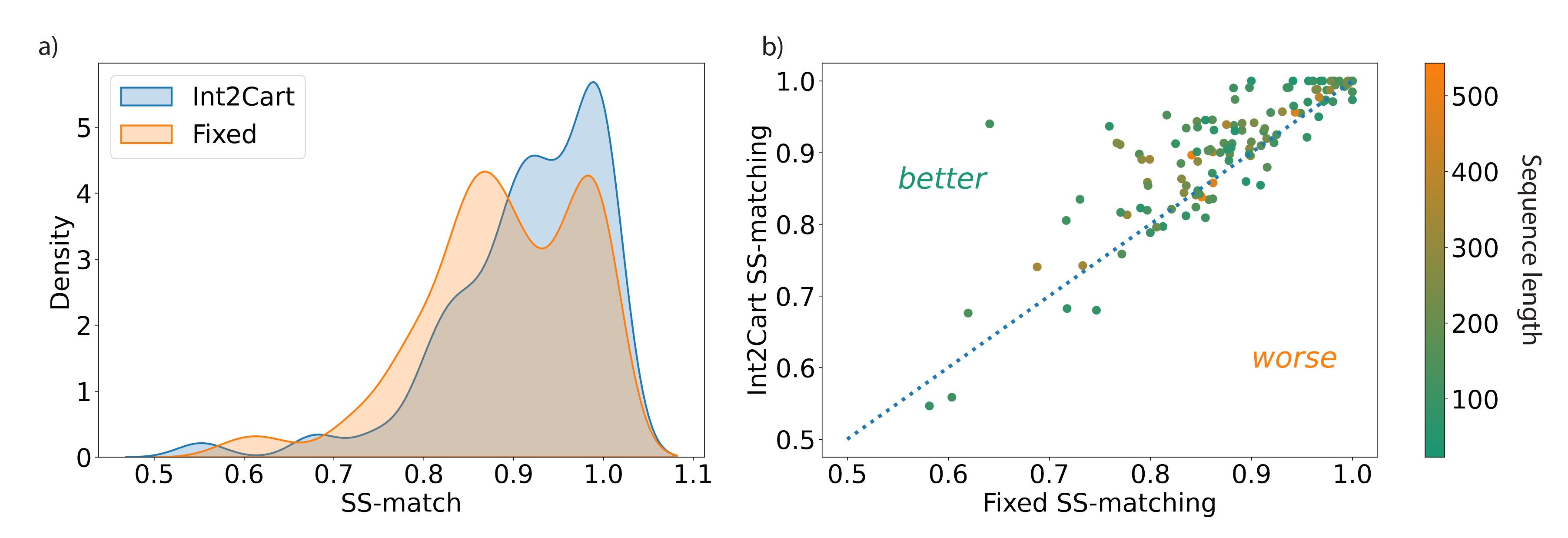}
\end{center}
\vspace{-4mm}
\caption{\textit{The accuracy of secondary structure assignment when internal coordinates are back-transformed to Cartesian coordinates using Int2Cart vs. Fixed.} (a) The SS-match distribution is calculated as the proportion of the 3-category secondary structure DSSP assignments (helix, strand, and coil)\cite{dssp} for each residue that matches the reference structure. (b) Comparison of SS-match for Int2Cart vs. Fixed  across the test set.}
\label{fig:newfig}
\end{figure}

\textit{Ablation studies}. To understand the importance of various inputs for prediction accuracy of Int2Cart and the importance of the bond length and bond angle targets in reconstructing Cartesian structures, we performed an ablation study by training separate deep learning models using subsets of the inputs, and reconstructing structures using only predicted bond lengths, only predicted bond angles, or using both. We have also trained models with additional inputs of $\chi_1$ torsion angles, along with $r_1$ and $\alpha_1$ sidechain bond lengths and bond angles as additional outputs, to evaluate how including sidechain information could improve prediction and reconstruction of backbone structures.  All ablation trials are reported in Table \ref{tab:ablation_input}. 

\begin{table}[ht]
\centering
\caption{\textit{Ablation studies of internal coordinate inputs}. Prediction RMSE and Cartesian structure reconstruction RMSD using predicted bond lengths and bond angles from an Int2Cart model taking different internal coordinate inputs. Each ablation on the input is repeated 3 times with different initializations of the machine learning model to obtain statistically meaningful results.}
\begin{tabular}{m{0.295\linewidth}m{0.175\linewidth}m{0.175\linewidth}m{0.275\linewidth}}
\hline\hline

\textbf{\small Model inputs} & \textbf{\small <d> RMSE (\AA)} & \textbf{\small <$\theta$> RMSE (\degree)} & \textbf{\small Cartesian RMSD (\AA)} \\ \hline

\small Residue type & \small 0.010$\pm$5E-5   & \small 1.86$\pm$0.002 & \small 4.97$\pm$ 0.01 \\
\small $\phi+\psi$ & \small 0.010$\pm$5E-5   & \small 1.67$\pm$0.004 & \small 4.24$\pm$0.02\\ 
\small $\phi+\psi$ + Residue type & \small 0.010$\pm$6E-5 & \small 1.63$\pm$0.004 & \small 4.05$\pm$0.02 \\  
\small $\phi+\psi + \omega$ + Residue type & \small 0.010$\pm$5E-5 & \small 1.48$\pm$0.006 & \small 3.56$\pm$0.03 (2.14 for 100 aa)\\
\small $\phi+\psi + \omega + \chi_1$ + Residue type & \small 0.009$\pm$2E-5 & \small 1.36$\pm$0.001 & \small 3.05$\pm$0.01\\
\hline
\end{tabular}
\label{tab:ablation_input}
\end{table}

The differences in predictions of the backbone bond lengths from different deep learning models are not significant, but prediction accuracy for backbone bond angles RMSE and reconstructed Cartesian structure RMSD are highly dependent on what information is available to the model. Specifically, a machine learning model that only knows about the residue types performs the worst with ~5 \r{A} in the reconstruction RMSD. Unsurprisingly based on statistical analysis of the PDB, backbone $\phi$ and $\psi$ torsion angles provide more information than residue types alone, and allows the reconstruction RMSD to decrease to 4.24 \r{A} on average. Including both $\phi$, $\psi$ and residue types further decreases the average reconstruction error to 4.05 \AA. As expected from the correlation of bond lengths and bond angles with $\omega$ torsion angles as well, including exact values for $\omega$ torsion angles also significantly improves the model and allows the reconstructed structure RMSD to decrease further to 3.56 \r{A} across the whole test set, and to 2.14 \r{A} for proteins normalized to 100 amino acids. When we tested the inclusion of sidechain $\chi_1$ torsion angles, we find that the 3D reconstruction model is even better, achieving an average reconstruction structure RMSD of 3.05 \r{A}. That is probably because the $\chi_1$ torsion angles are indicative of avoidable steric clashes between protein backbones and side chains to create more accurate descriptions of subsequent backbone bond geometries, even though side chain atoms are not explicitly treated during structure reconstruction. 

\begin{table}[h]
\centering
\caption{\textit{Ablation studies of Cartesian coordinate reconstructions}. Cartesian structure reconstruction RMSD using fixed bond lengths and angles, using predicted bond lengths and fixed angles, fixed bond lengths and predicted bond angles, and both predicted bond lengths and bond angles.}
\begin{tabular}{l|c}
\hline \hline
\textbf{\small Source of bond lengths \& bond angles} & \textbf{\small Reconstructed structure RMSD (\AA)} \\
\hline
\small Fixed bond lengths and angles &  \small 5.07$\pm$3.47\\
\small Predicted bond lengths and fixed bond angles & \small 5.06$\pm$3.46 \\
\small Fixed bond lengths and predicted bond angles & \small 3.51$\pm$2.65\\
\small Predicted bond lengths and bond angles & \small 3.51$\pm$2.65\\
\hline
\end{tabular}
\label{tab:ablation_output}
\end{table}

Table \ref{tab:ablation_output} shows the Cartesian structure RMSDs for protein structures reconstructed with predicted bond lengths, predicted bond angles and both, and compared with using fixed bond lengths and bond angles. Interestingly, the reconstruction quality does not depend on the direct prediction of bond lengths, as it essentially has no effect on the reconstructed structures, which my have been anticipated from the fact that bond length errors are on par with the variance. But this final ablation study provides direct evidence that accurate predictions of bond angles are of primary importance for the quality of the reconstructed Cartesian structures.

\textit{Evaluation of the method on more challenging test datasets.} Finally we consider two independent test data sets that are quite different from the originally defined test set from SidechainNet, performed to evaluate the transferability of the Int2Cart model. Table \ref{tab:reconst_accuracy} reports the prediction RMSE for all bond length and bond angles targets of the CASP-12 protein data set and the mean Cartesian reconstruction error. Compared with the previous test dataset results, the predictions on $\theta_3$ becomes slightly worse in terms of RMSE and correlation coefficients, but all other prediction targets are very close to the original results described in Table 1. The reconstructed structure RMSD on the CASP-12 test dataset is on average 4.5 \r{A}, but this shift to higher RMSD values relative to the original test set is due to the higher proportion of longer proteins. This is supported by the fact that the reconstructed RMSD as a function of sequence length for the CASP-12 data are indistinguishable from the primary test dataset as illustrated in Supplementary Figure 10. Additionally, the reconstructed RMSD for proteins normalized to 100 amino acids in the CASP12 test set is 2.06 \r{A}, in close agreement with the results of the original test set. 

\begin{table}[ht]
    \centering
\caption{\textit{Int2Cart prediction accuracy and reconstructed Cartesian structure evaluated on the more challenging CASP12 dataset}. Accuracy is assessed in terms of root-mean-square error (RMSE) and Pearson correlation coefficients ($R$) for bond and angle predictions, and Cartesian reconstruction evaluated as root-mean-square-deviation (RMSD) with respect to the reference PDB structure.}
\begin{tabular}{lcc}
\hline\hline
\textbf{\small Data type}& \small RMSE & \small $R$     \\ \hline
\textbf{\small N-C$_\alpha$-C  ($\theta_1$)} & \small 1.79\degree & \small 0.69 \\
\textbf{C$_\alpha$-C-N ($\theta_2$)} & 1.01\degree & 0.60\\
\textbf{C-N-C$_\alpha$  ($\theta_3$)} & 1.71\degree & 0.41\\
                           
\hline
\textbf{N-C$_\alpha$ ($d_1$)} & 0.009 \AA  & 0.39 \\ 
\textbf{C$_\alpha$-C ($d_2$)} & 0.010 \AA  & 0.48  \\
\textbf{C-N ($d_3$)} & 0.011 \AA & 0.39 \\  
\hline\hline
\textbf{\small Reconstruction}& RMSD \\
{\small all CASP12 proteins} & 4.49 \AA  &  \\ 
{\small CASP12 proteins (normalized to 100 aa)} & 2.06 \AA  &  \\ 
\hline\hline
\end{tabular}

\label{tab:reconst_accuracy}
\end{table}

\begin{figure}[ht]
\begin{center}
\includegraphics[width=0.7\textwidth]{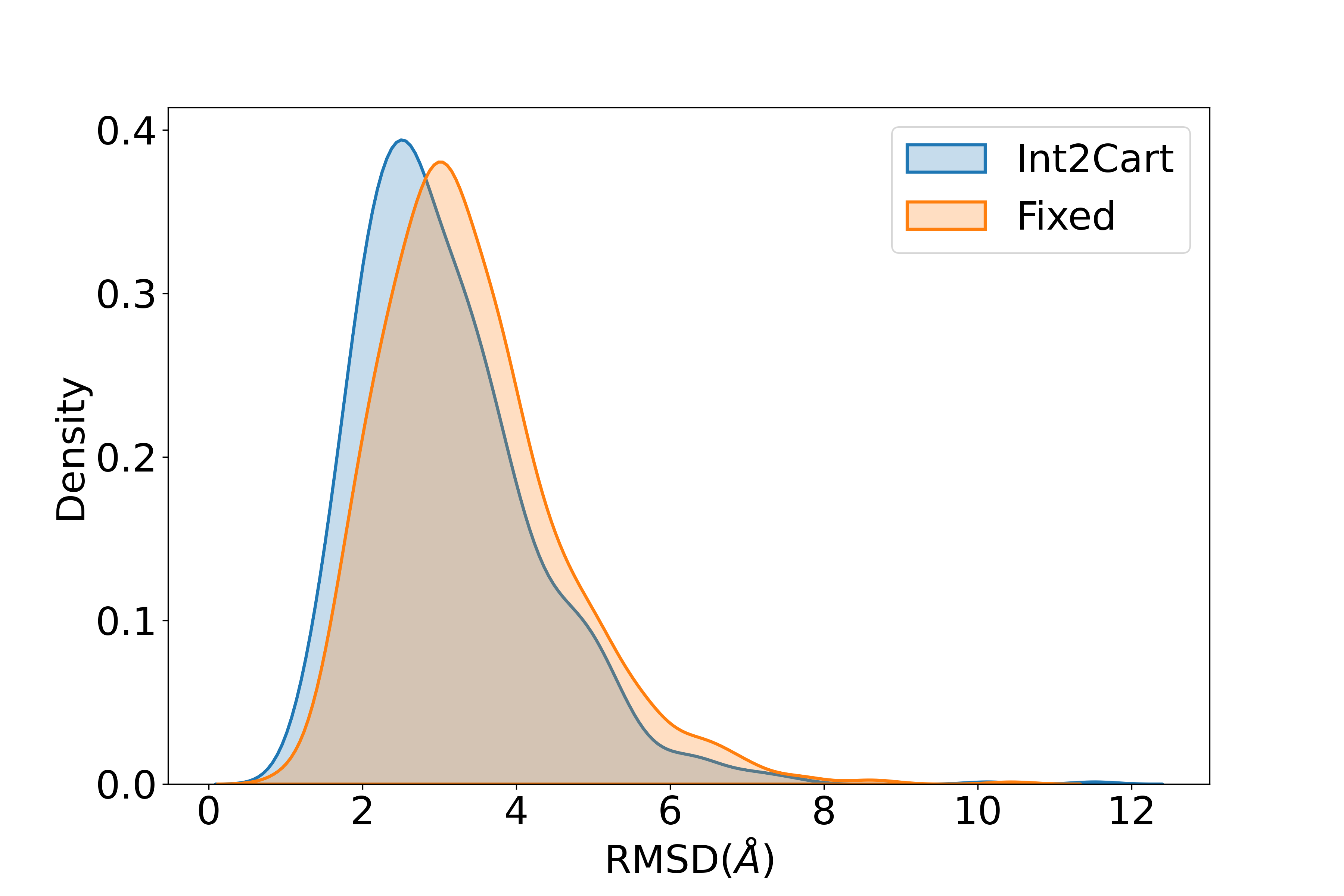}
\end{center}
\vspace{-5mm}
\caption{\textit{Comparison of distribution of reconstruction RMSD for individual conformaions in the Sic1 IDP ensemble.} Structures reconstructed with Int2Cart method on average has lower RMSD to their original structures compared with using fixed bond lengths and bond angles.}
\label{fig:idp}
\end{figure}

In the second stress test we show that our Int2Cart method can improve upon the Cartesian reconstruction of an ensemble of structures of a disordered protein compared to Fixed bond lengths and angles. Figure \ref{fig:idp} compares the Cartesian reconstruction RMSD distributions for Int2Cart and 

Fixed for the Sic1 IDP ensemble, in which we find that the Int2Cart method is overall closer to the original ensemble, with a 3.1 \r{A} average RMSD compared to the Fixed method that has a mean RMSD of 3.4 \r{A}. We have also checked the number of steric clashes in the structures generated from these two methods. A steric clash is defined as two atoms in the structure that are closer to 0.6 times the sum of the van der Waals radii of the two atoms.\cite{mcsce} Out of the 1000 conformations, 73 structures generated from Int2Cart contained steric clashes, which means 92.7\% of the  structures are clash-free. By comparison, 102 structures generated using fixed bond lengths and bond angles contained steric clashes, which translates to 89.8\% of clash-free structures. A higher proportion of clash-free structures is meaningful because typically structures containing clashes are discarded, and a method with higher proportion of clash-free structures wastes less computational resources, and supports the application of the Int2Cart algorithm to disordered protein ensembles.

\section*{\fontsize{12}{12}\selectfont DISCUSSION AND CONCLUSION}
In this work we have developed a new machine learning approach to the generic representation problem of internal coordinates (bond lengths, valence angles, and dihedral angles) and how to increase the fidelity of the back-transformation to 3D Cartesian coordinates. The Int2Cart algorithm utilizes a gated recurrent unit  neural network to predict real-valued backbone bond lengths and bond angles for each residue of a complete protein sequence given its torsion angle profile. In summary Int2Cart can reconstruct the Cartesian structure of 95\% of test proteins with RMSDs that are significant improvements over the fixed backbone bond lengths and bond angles that are the standard practice in a large variety of protein modelling approaches. The success of our algorithm on the more dissimilar CASP-12 prediction targets and across IDP ensembles further validates that the Int2Cart algorithm is transferable among different types of proteins, and can consistently improve the quality of Cartesian structure reconstruction. 

In its current form the Int2Cart algorithm only generates backbone structures for the target proteins, although we can improve Cartesian reconstruction performance with the inclusion the $\chi_1$ torsion and predicting $r_1$ and $\alpha_1$. Theoretical approaches such as the Monte Carlo Side Chain Ensemble (MC-SCE) approach can utilize the backbone from Int2Cart to calculate side chain ensembles in order to complete the full structure.\cite{Bhowmick2015} It is also clear that there is still room for improvement in the Cartesian reconstruction of larger proteins, and the inherent scaling of error with respect to sequence length is inevitable for a deep learning model that predicts internal coordinates in a sequential manner (i.e. a GRU model). Therefore, it may be possible to improve the quality of Cartesian structure reconstruction with a distance-based neural network model, i.e., by representing the 3D coordinates of the structure directly. 

Nevertheless, the model in its current form already provides a useful computational tool to greatly improve the quality of protein structures reconstructed from backbone torsion angles only, regardless of globular folded proteins or disordered protein ensembles. We envision Int2Cart should see broad use in structure refinement and validation\cite{wilson1998checks,kleywegt2009vital} and development of protein force fields that could benefit from more accurate valence models of backbone bond lengths and bond angles conditioned on other geometrical or sequence features.\cite{conway2014relaxation}  Finally, the Int2Cart GRU neural network model could also be useful for other chain molecules, only requiring retraining with new data if available for systems such as nucleic acids and lipids. 

\section*{\fontsize{12}{12}\selectfont 
AUTHOR CONTRIBUTIONS}
\vspace{-3mm}
J.L. and T.H.-G. designed the project. J.L. designed and wrote the Int2Cart software. O.Z. also provided input on the neural network design and tuning of the model and valuable critiques including testing. J.L. and T.H.-G. wrote the paper and all authors provided valuable input and discussion.

\section*{\fontsize{12}{12}\selectfont ACKNOWLEDGEMENTS}
\vspace{-3mm}
T.H.-G. and J.D.F.-K. acknowledge funding from the National Institute of Health under Grant 5R01GM127627-04. J.D.F.-K. also acknowledges support from the Natural Sciences and Engineering Re-search Council of Canada (2016-06718) and from the Canada Research Chairs Program.

\bibliographystyle{unsrtnat}
\bibliography{references}

\begin{thebibliography}{32}
\providecommand{\natexlab}[1]{#1}
\providecommand{\url}[1]{\texttt{#1}}
\expandafter\ifx\csname urlstyle\endcsname\relax
  \providecommand{\doi}[1]{doi: #1}\else
  \providecommand{\doi}{doi: \begingroup \urlstyle{rm}\Url}\fi

\bibitem[Baker et~al.(1999)Baker, Kinghorn, and Pulay]{Baker1999}
Jon Baker, Don Kinghorn, and Peter Pulay.
\newblock Geometry optimization in delocalized internal coordinates: An
  efficient quadratically scaling algorithm for large molecules.
\newblock \emph{The Journal of Chemical Physics}, 110\penalty0 (11):\penalty0
  4986--4991, 1999.
\newblock \doi{10.1063/1.478397}.
\newblock URL \url{https://doi.org/10.1063/1.478397}.

\bibitem[Schwieters and Clore(2001)]{SCHWIETERS2001}
Charles~D. Schwieters and G.Marius Clore.
\newblock Internal coordinates for molecular dynamics and minimization in
  structure determination and refinement.
\newblock \emph{Journal of Magnetic Resonance}, 152\penalty0 (2):\penalty0
  288--302, 2001.
\newblock ISSN 1090-7807.
\newblock \doi{https://doi.org/10.1006/jmre.2001.2413}.
\newblock URL
  \url{https://www.sciencedirect.com/science/article/pii/S1090780701924139}.

\bibitem[Adcock and McCammon(2006)]{Adcock2006}
Stewart~A. Adcock and J.~Andrew McCammon.
\newblock Molecular dynamics: Survey of methods for simulating the activity of
  proteins.
\newblock \emph{Chemical Reviews}, 106\penalty0 (5):\penalty0 1589--1615, 2006.
\newblock \doi{10.1021/cr040426m}.
\newblock URL \url{https://doi.org/10.1021/cr040426m}.
\newblock PMID: 16683746.

\bibitem[wwPDB consortium(2018)]{PDB2018}
wwPDB consortium.
\newblock {Protein Data Bank: the single global archive for 3D macromolecular
  structure data}.
\newblock \emph{Nucleic Acids Research}, 47\penalty0 (D1):\penalty0 D520--D528,
  10 2018.
\newblock ISSN 0305-1048.
\newblock \doi{10.1093/nar/gky949}.
\newblock URL \url{https://doi.org/10.1093/nar/gky949}.

\bibitem[Parsons et~al.(2005)Parsons, Holmes, Rojas, Tsai, and Strauss]{nerf}
Jerod Parsons, J~Bradley Holmes, J~Maurice Rojas, Jerry Tsai, and Charlie~EM
  Strauss.
\newblock Practical conversion from torsion space to cartesian space for in
  silico protein synthesis.
\newblock \emph{Journal of computational chemistry}, 26\penalty0 (10):\penalty0
  1063--1068, 2005.

\bibitem[Craveur et~al.(2013)Craveur, Joseph, Poulain, de~Brevern, and
  Rebehmed]{craveur2013cis}
Pierrick Craveur, Agnel~Praveen Joseph, Pierre Poulain, Alexandre~G de~Brevern,
  and Joseph Rebehmed.
\newblock Cis--trans isomerization of omega dihedrals in proteins.
\newblock \emph{Amino acids}, 45\penalty0 (2):\penalty0 279--289, 2013.

\bibitem[Holmes and Tsai(2004)]{holmes2004some}
J~Bradley Holmes and Jerry Tsai.
\newblock Some fundamental aspects of building protein structures from fragment
  libraries.
\newblock \emph{Protein science}, 13\penalty0 (6):\penalty0 1636--1650, 2004.

\bibitem[Ramachandran et~al.(1963)Ramachandran, Ramakrishnan, and
  Sasisekharan]{RAMACHANDRAN196395}
G.N. Ramachandran, C.~Ramakrishnan, and V.~Sasisekharan.
\newblock Stereochemistry of polypeptide chain configurations.
\newblock \emph{Journal of Molecular Biology}, 7\penalty0 (1):\penalty0 95--99,
  1963.
\newblock ISSN 0022-2836.
\newblock \doi{https://doi.org/10.1016/S0022-2836(63)80023-6}.
\newblock URL
  \url{https://www.sciencedirect.com/science/article/pii/S0022283663800236}.

\bibitem[Shapovalov and Dunbrack~Jr(2011)]{dumbrack_library}
Maxim~V Shapovalov and Roland~L Dunbrack~Jr.
\newblock A smoothed backbone-dependent rotamer library for proteins derived
  from adaptive kernel density estimates and regressions.
\newblock \emph{Structure}, 19\penalty0 (6):\penalty0 844--858, 2011.

\bibitem[Karplus(1996)]{karplus1996experimentally}
P~Andrew Karplus.
\newblock Experimentally observed conformation-dependent geometry and hidden
  strain in proteins.
\newblock \emph{Protein Science}, 5\penalty0 (7):\penalty0 1406--1420, 1996.

\bibitem[Jiang et~al.(1995)Jiang, Cao, Teppen, Newton, and
  Schaefer]{jiang1995predictions}
Xiaoqin Jiang, Ming Cao, Brian Teppen, Susan~Q Newton, and Lothar Schaefer.
\newblock Predictions of protein backbone structural parameters from first
  principles: Systematic comparisons of calculated nc (. alpha.)-c'angles with
  high-resolution protein crystallographic results.
\newblock \emph{The Journal of Physical Chemistry}, 99\penalty0 (26):\penalty0
  10521--10525, 1995.

\bibitem[Sch{\'a}fer et~al.(1995)Sch{\'a}fer, Cao, and
  Meadows]{schafer1995predictions}
Lothar Sch{\'a}fer, Ming Cao, and Mary~Jane Meadows.
\newblock Predictions of protein backbone bond distances and angles from first
  principles.
\newblock \emph{Biopolymers: Original Research on Biomolecules}, 35\penalty0
  (6):\penalty0 603--606, 1995.

\bibitem[Yu et~al.(2001)Yu, Norman, Sch{\"a}fer, Ramek, Peeters, and
  Van~Alsenoy]{yu2001ab}
Ching-Hsing Yu, Mya~A Norman, Lothar Sch{\"a}fer, Michael Ramek, Anik Peeters,
  and Christian Van~Alsenoy.
\newblock Ab initio conformational analysis of n-formyl l-alanine amide
  including electron correlation.
\newblock \emph{Journal of Molecular Structure}, 567:\penalty0 361--374, 2001.

\bibitem[Berkholz et~al.(2009)Berkholz, Shapovalov, Dunbrack~Jr, and
  Karplus]{berkholz2009conformation}
Donald~S Berkholz, Maxim~V Shapovalov, Roland~L Dunbrack~Jr, and P~Andrew
  Karplus.
\newblock Conformation dependence of backbone geometry in proteins.
\newblock \emph{Structure}, 17\penalty0 (10):\penalty0 1316--1325, 2009.

\bibitem[Improta et~al.(2015{\natexlab{a}})Improta, Vitagliano, and
  Esposito]{improta2015determinants}
Roberto Improta, Luigi Vitagliano, and Luciana Esposito.
\newblock The determinants of bond angle variability in protein/peptide
  backbones: A comprehensive statistical/quantum mechanics analysis.
\newblock \emph{Proteins: Structure, Function, and Bioinformatics}, 83\penalty0
  (11):\penalty0 1973--1986, 2015{\natexlab{a}}.

\bibitem[Improta et~al.(2015{\natexlab{b}})Improta, Vitagliano, and
  Esposito]{improta2015bond}
Roberto Improta, Luigi Vitagliano, and Luciana Esposito.
\newblock Bond distances in polypeptide backbones depend on the local
  conformation.
\newblock \emph{Acta Crystallographica Section D: Biological Crystallography},
  71\penalty0 (6):\penalty0 1272--1283, 2015{\natexlab{b}}.

\bibitem[Lundgren and Niemi(2012)]{lundgren2012correlation}
Martin Lundgren and Antti~J Niemi.
\newblock Correlation between protein secondary structure, backbone bond
  angles, and side-chain orientations.
\newblock \emph{Physical Review E}, 86\penalty0 (2):\penalty0 021904, 2012.

\bibitem[Ravikumar et~al.(2019)Ravikumar, Ramakrishnan, and
  Srinivasan]{ravikumar2019stereochemical}
Ashraya Ravikumar, Chandrasekharan Ramakrishnan, and Narayanaswamy Srinivasan.
\newblock Stereochemical assessment of ($\varphi$, $\psi$) outliers in protein
  structures using bond geometry-specific ramachandran steric-maps.
\newblock \emph{Structure}, 27\penalty0 (12):\penalty0 1875--1884, 2019.

\bibitem[King and Koes(2021)]{King2021}
J.~E. King and D.~R. Koes.
\newblock Sidechainnet: An all-atom protein structure dataset for machine
  learning.
\newblock \emph{Proteins}, 89\penalty0 (11):\penalty0 1489--1496, 2021.
\newblock ISSN 0887-3585 (Print) 0887-3585.
\newblock \doi{10.1002/prot.26169}.

\bibitem[AlQuraishi(2019)]{AlQuraishi2019}
Mohammed AlQuraishi.
\newblock Proteinnet: a standardized data set for machine learning of protein
  structure.
\newblock \emph{BMC Bioinformatics}, 20\penalty0 (1):\penalty0 311, 2019.
\newblock ISSN 1471-2105.
\newblock \doi{10.1186/s12859-019-2932-0}.
\newblock URL \url{https://doi.org/10.1186/s12859-019-2932-0}.

\bibitem[Mittag et~al.(2008)Mittag, Orlicky, Choy, Tang, Lin, Sicheri, Kay,
  Tyers, and Forman-Kay]{Mittag2008}
Tanja Mittag, Stephen Orlicky, Wing-Yiu Choy, Xiaojing Tang, Hong Lin, Frank
  Sicheri, Lewis~E. Kay, Mike Tyers, and Julie~D. Forman-Kay.
\newblock Dynamic equilibrium engagement of a polyvalent ligand with a
  single-site receptor.
\newblock \emph{Proceedings of the National Academy of Sciences}, 105\penalty0
  (46):\penalty0 17772--17777, 2008.
\newblock \doi{10.1073/pnas.0809222105}.
\newblock URL \url{https://www.pnas.org/doi/abs/10.1073/pnas.0809222105}.

\bibitem[Gomes et~al.(2020)Gomes, Krzeminski, Namini, Martin, Mittag,
  Head-Gordon, Forman-Kay, and Gradinaru]{Gomes2020}
Gregory-Neal~W. Gomes, Micka{\"e}l Krzeminski, Ashley Namini, Erik~W. Martin,
  Tanja Mittag, Teresa Head-Gordon, Julie~D. Forman-Kay, and Claudiu~C.
  Gradinaru.
\newblock Conformational ensembles of an intrinsically disordered protein
  consistent with nmr, saxs, and single-molecule fret.
\newblock \emph{Journal of the American Chemical Society}, 142\penalty0
  (37):\penalty0 15697--15710, 2020.
\newblock \doi{10.1021/jacs.0c02088}.
\newblock URL \url{https://doi.org/10.1021/jacs.0c02088}.
\newblock PMID: 32840111.

\bibitem[Cho et~al.(2014)Cho, Van~Merri{\"e}nboer, Bahdanau, and Bengio]{gru}
Kyunghyun Cho, Bart Van~Merri{\"e}nboer, Dzmitry Bahdanau, and Yoshua Bengio.
\newblock On the properties of neural machine translation: Encoder-decoder
  approaches.
\newblock \emph{arXiv preprint arXiv:1409.1259}, 2014.

\bibitem[Glorot et~al.(2011)Glorot, Bordes, and Bengio]{relu}
Xavier Glorot, Antoine Bordes, and Yoshua Bengio.
\newblock Deep sparse rectifier neural networks.
\newblock In \emph{Proceedings of the fourteenth international conference on
  artificial intelligence and statistics}, pages 315--323. JMLR Workshop and
  Conference Proceedings, 2011.

\bibitem[Kingma and Ba(2015)]{Kingma2015}
Diederik~P. Kingma and Jimmy Ba.
\newblock Adam: A method for stochastic optimization.
\newblock \emph{CoRR}, abs/1412.6980, 2015.

\bibitem[Tan et~al.(2016)Tan, Gu, Jedrzejczak, and Joachimiak]{5jmb_structure}
K.~Tan, M.~Gu, R.~Jedrzejczak, and A.~Joachimiak.
\newblock The crystal structure of the n-terminal domain of a novel cellulases
  from bacteroides coprocola.
\newblock \emph{PDB}, 2016.
\newblock URL \url{https://www.rcsb.org/structure/5JMB}.

\bibitem[Kabsch and Sander(1983)]{dssp}
Wolfgang Kabsch and Christian Sander.
\newblock Dictionary of protein secondary structure: pattern recognition of
  hydrogen-bonded and geometrical features.
\newblock \emph{Biopolymers: Original Research on Biomolecules}, 22\penalty0
  (12):\penalty0 2577--2637, 1983.

\bibitem[Bhowmick and Head-Gordon(2015{\natexlab{a}})]{mcsce}
Asmit Bhowmick and Teresa Head-Gordon.
\newblock A monte carlo method for generating side chain structural ensembles.
\newblock \emph{Structure}, 23\penalty0 (1):\penalty0 44--55,
  2015{\natexlab{a}}.

\bibitem[Bhowmick and Head-Gordon(2015{\natexlab{b}})]{Bhowmick2015}
Asmit Bhowmick and Teresa Head-Gordon.
\newblock A monte carlo method for generating side chain structural ensembles.
\newblock \emph{Structure}, 23\penalty0 (1):\penalty0 44--55,
  2015{\natexlab{b}}.
\newblock ISSN 0969-2126.
\newblock \doi{http://dx.doi.org/10.1016/j.str.2014.10.011}.
\newblock URL
  \url{http://www.sciencedirect.com/science/article/pii/S0969212614003566}.

\bibitem[Wilson et~al.(1998)Wilson, Butterworth, Dauter, Lamzin, Walsh, Wodak,
  Pontius, Richelle, Vaguine, Sander, et~al.]{wilson1998checks}
KS~Wilson, S~Butterworth, Z~Dauter, VS~Lamzin, M~Walsh, S~Wodak, J~Pontius,
  J~Richelle, A~Vaguine, C~Sander, et~al.
\newblock Who checks the checkers? four validation tools applied to eight
  atomic resolution structures.
\newblock \emph{Journal of Molecular Biology}, 276\penalty0 (2):\penalty0 417,
  1998.

\bibitem[Kleywegt(2009)]{kleywegt2009vital}
Gerard~J Kleywegt.
\newblock On vital aid: the why, what and how of validation.
\newblock \emph{Acta Crystallographica Section D: Biological Crystallography},
  65\penalty0 (2):\penalty0 134--139, 2009.

\bibitem[Conway et~al.(2014)Conway, Tyka, DiMaio, Konerding, and
  Baker]{conway2014relaxation}
Patrick Conway, Michael~D Tyka, Frank DiMaio, David~E Konerding, and David
  Baker.
\newblock Relaxation of backbone bond geometry improves protein energy
  landscape modeling.
\newblock \emph{Protein Science}, 23\penalty0 (1):\penalty0 47--55, 2014.

\end{thebibliography}
\end{document}


\title{\textbf{Supporting Information\\ Learning Correlations between Internal Coordinates to improve 3D Cartesian Coordinates for Proteins}}
\date{}
\author{Jie Li$^{1}$, Oufan Zhang$^{1}$, Seokyoung Lee$^{1}$, Ashley Namini$^{2}$, Zi Hao Liu$^{2,3}$, \\
João Miguel Correia Teixeira$^{2}$, Julie D Forman-Kay$^{2,3}$, Teresa Head-Gordon$^{1,4,5}$}
\maketitle
\noindent
$^1$Kenneth S. Pitzer Theory Center and Department of Chemistry, University of California, Berkeley, CA, USA\\
$^2$Molecular Medicine Program, Hospital for Sick Children, Toronto, Ontario M5S 1A8, Canada\\
$^3$Department of Biochemistry, University of Toronto, Toronto, Ontario M5G 1X8, Canada\\
$^4$Chemical Sciences Division, Lawrence Berkeley National Laboratory, Berkeley, CA, USA\\
$^5$Departments of Bioengineering and Chemical and Biomolecular Engineering, University of California, Berkeley, CA, USA\\
\vspace{-5mm}
\begin{center}
corresponding author: thg@berkeley.edu
\end{center}

\begin{table}[h]
    \centering
    \caption{Mean and standard deviations used to rescale model predictions for bond lengths (\AA) and bond angles (\textbf{rad})}
    \begin{tabular}{lcc}
        \hline\hline

            \textbf{Data type}& Mean & Standard deviation     \\ \hline
        \textbf{N-C$_\alpha$ bond length} & 1.460  & 0.0118 \\ 
        \textbf{C$_\alpha$-C bond length} & 1.525 & 0.0123  \\
        \textbf{C-N bond length} & 1.331& 0.0095 \\  \hline
        \textbf{N-C$_\alpha$-C bond angle} & 1.941 & 0.0472 \\
        \textbf{C$_\alpha$-C-N bond angle} & 2.034 & 0.0413\\
        \textbf{C-N-C$_\alpha$ bond angle} & 2.122 & 0.0480\\
                                       
        \hline
    \end{tabular}

\label{tab:meanstd}
\end{table}

\newpage

\begin{figure}[H]
\begin{center}
\includegraphics[width=0.98\textwidth]{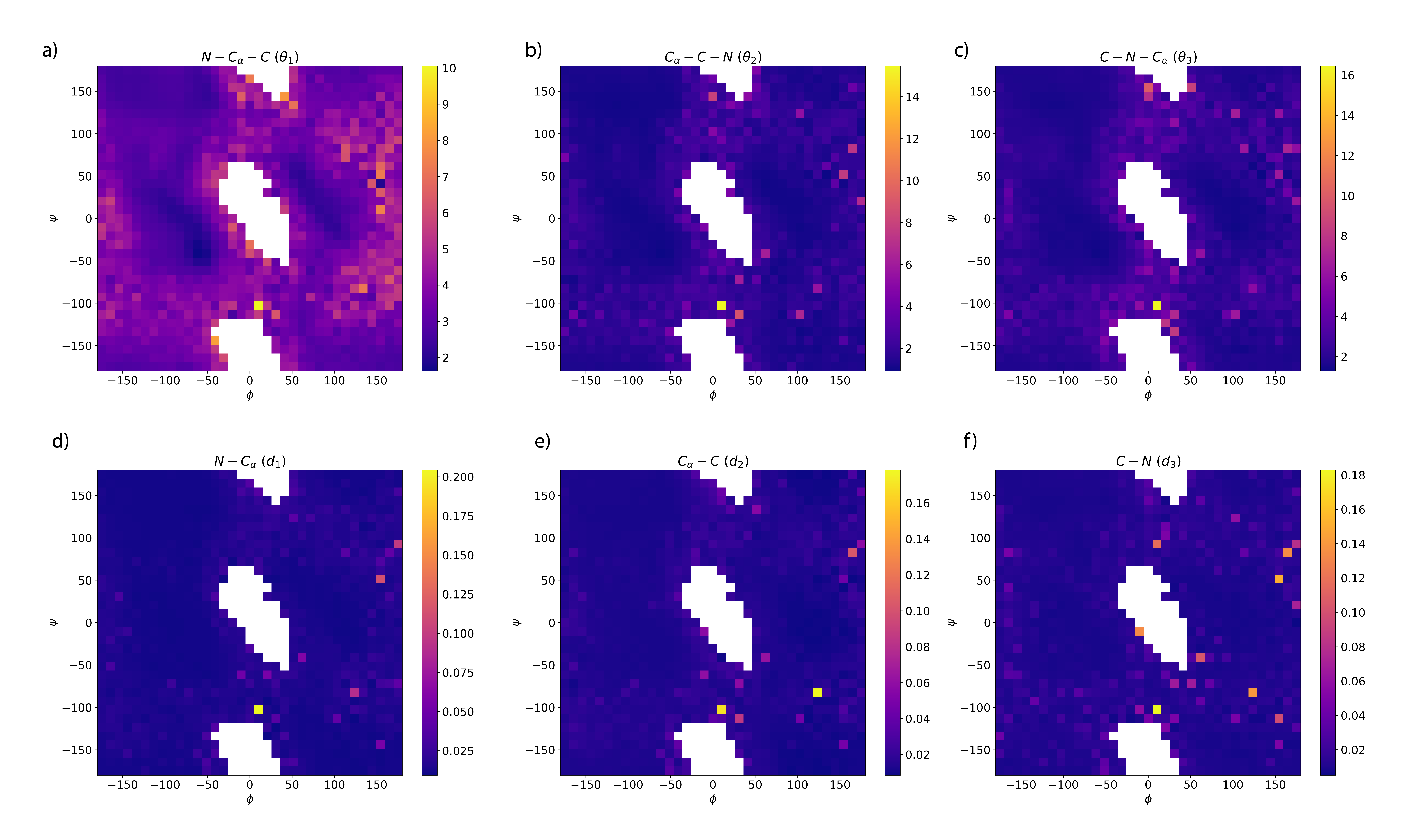}
\end{center}
\caption{\textit{Variations in the standard deviation (STD) of bond angle and bond lengths as a function of $\phi$,$\psi$.} The regions of red correspond to larger STD while the region in blue have much smaller STD.}
\label{fig:dist-std}
\end{figure}

\begin{figure}[H]
\begin{center}
\includegraphics[width=0.98\textwidth]{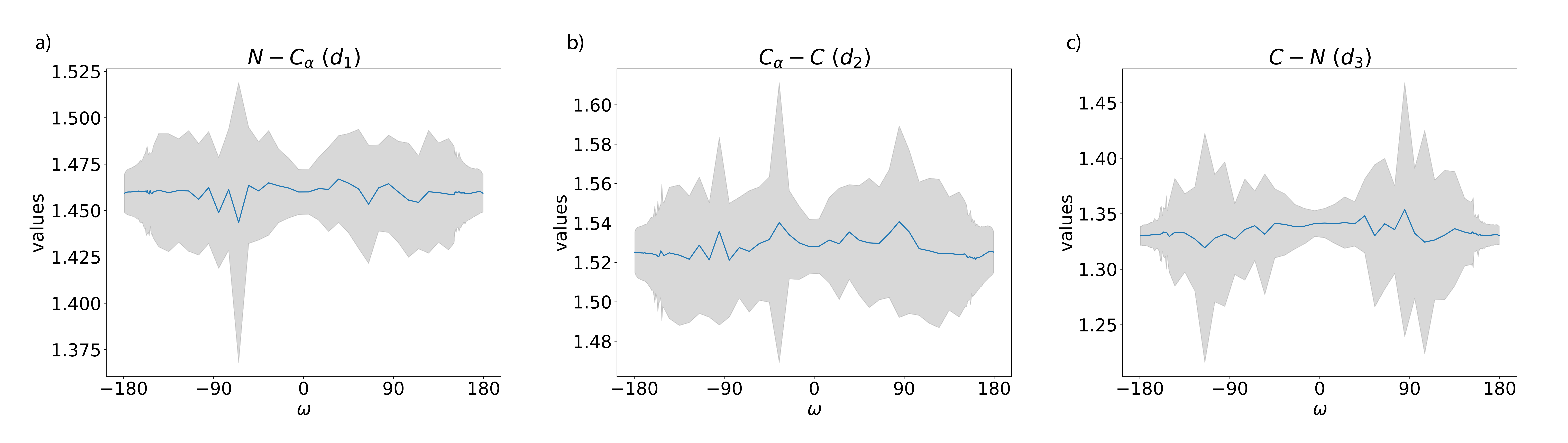}
\end{center}
\caption{\textit{Mean values and standard deviations of bond lengths as a function of $\omega$.} The blue solid lines represent mean values of bond lengths at specific $\omega$ torsion angles, and the gray regions correspond to one standard deviation.}
\label{fig:dist-std}
\end{figure}

\begin{figure}[H]
\begin{center}
\includegraphics[width=0.98\textwidth]{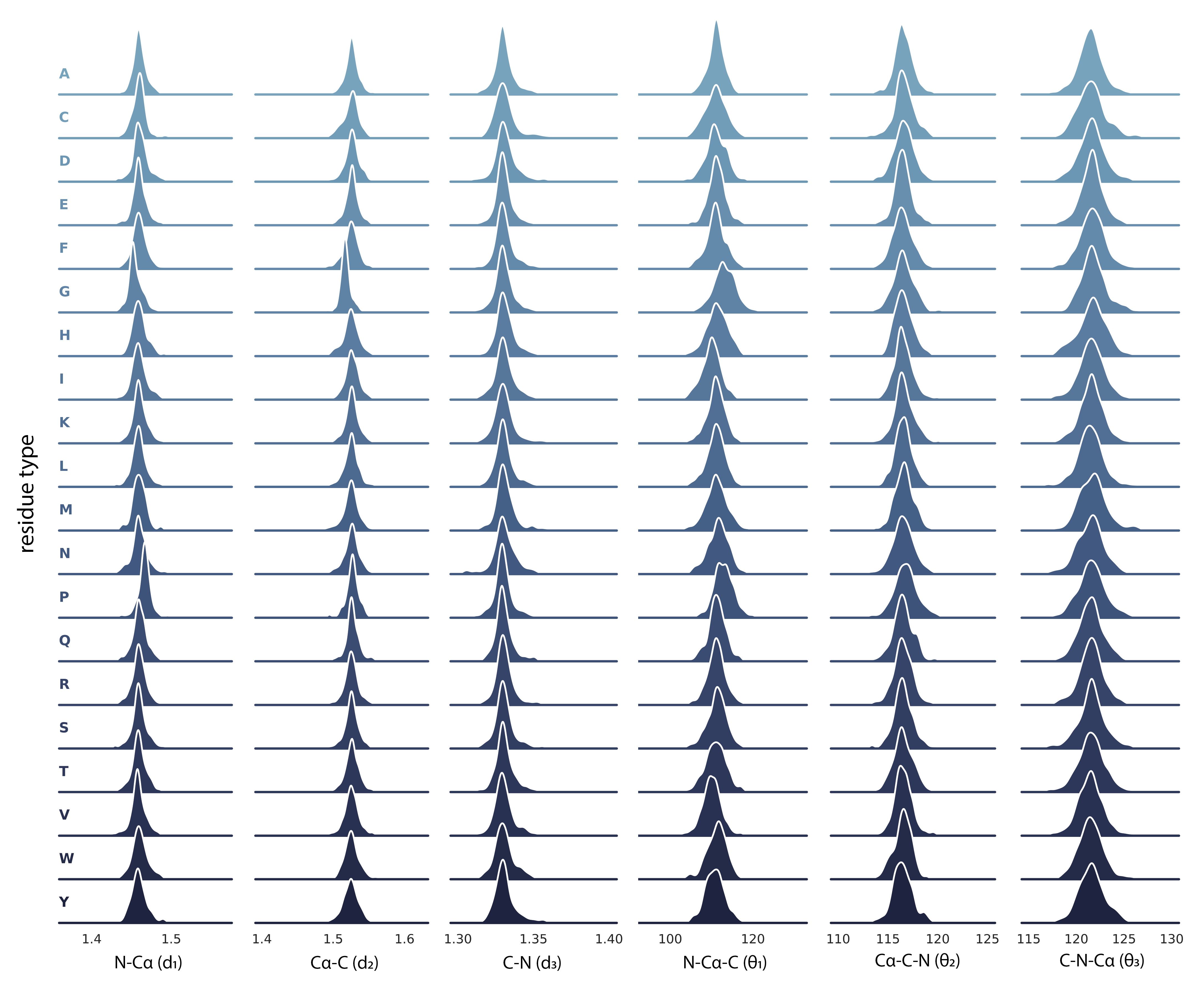}
\end{center}
\caption{\textit{Distributions of bond lengths and bond angles as a function of residue type.} Shown for all twenty amino acids.}
\label{fig:dist-std}
\end{figure}
\begin{figure}
\begin{center}
\includegraphics[width=0.98\textwidth]{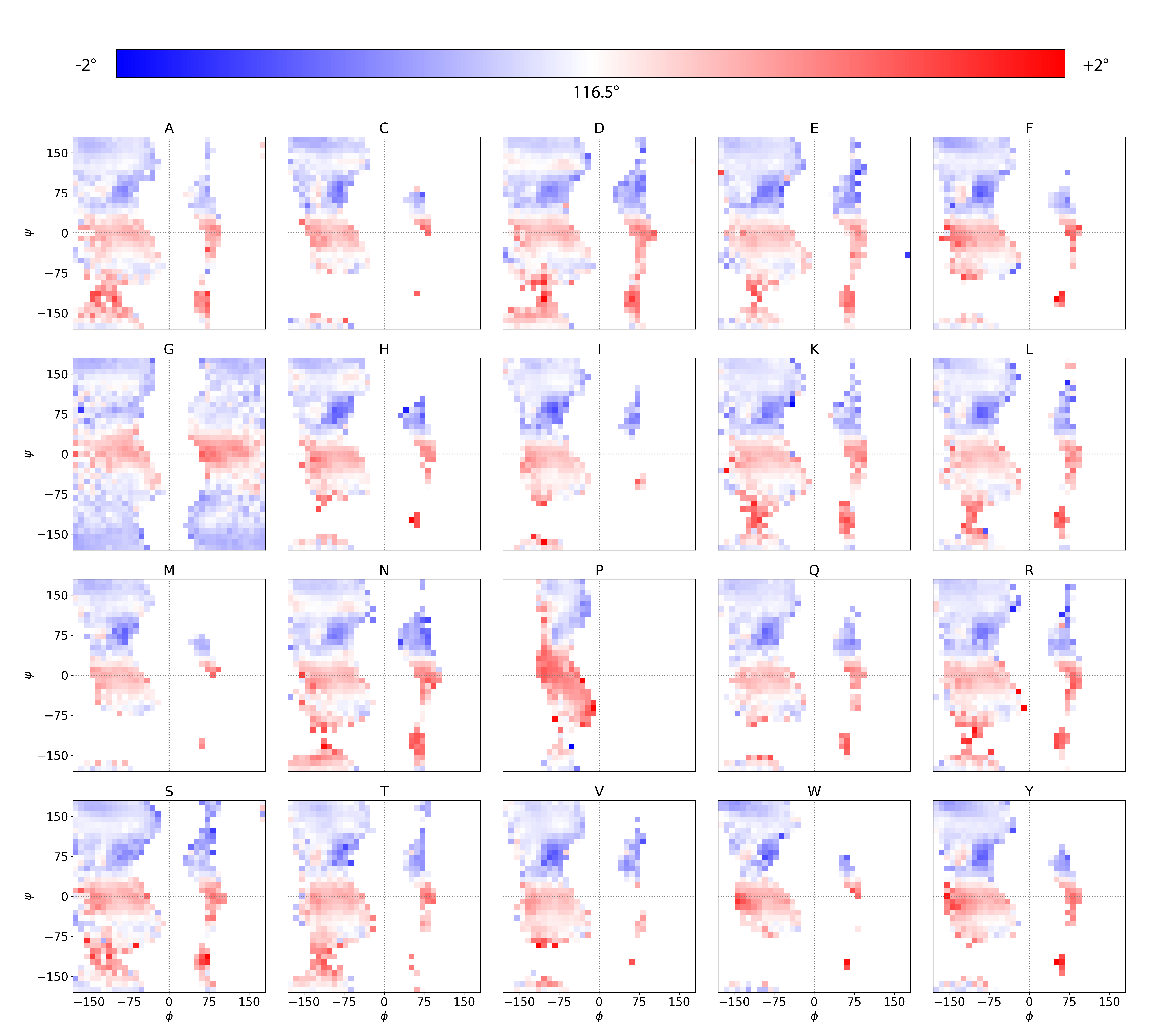}
\end{center}
\caption{\textit{$C_{\alpha}-C-N$ bond angle deviations from the mean values averaged over $\phi$ and $\psi$ angles as a function of residue type}. The regions of red correspond to longer bonds while the region in blue show reduced bond values relative to the mean. The $C_{\alpha}-C-N$ bond angles were categorized according to $\phi$ and $\psi$ angles rounded to the closest tens.}
\label{fig:theta2-res}
\end{figure}

\begin{figure}
\begin{center}
\includegraphics[width=0.98\textwidth]{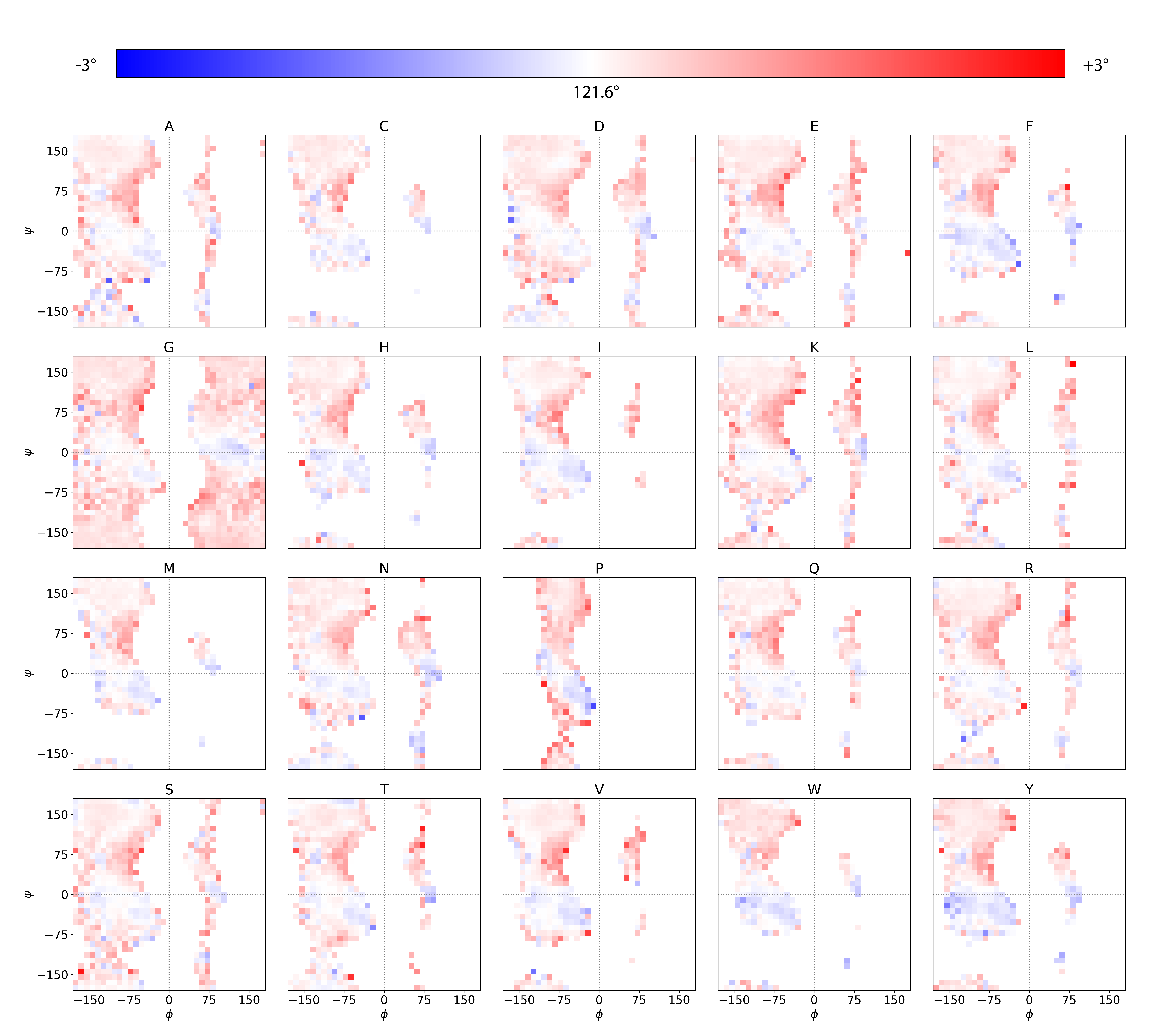}
\end{center}
\caption{\textit{$C-N-C_{\alpha}$ bond angle deviations from the mean values averaged over $\phi$ and $\psi$ angles as a function of residue type}. The regions of red correspond to longer bonds while the region in blue show reduced bond values relative to the mean. The $C-N-C_{\alpha}$ bond angles were categorized according to $\phi$ and $\psi$ angles rounded to the closest tens.}
\label{fig:theta3-res}
\end{figure}

\begin{figure}
\begin{center}
\includegraphics[width=0.9\textwidth]{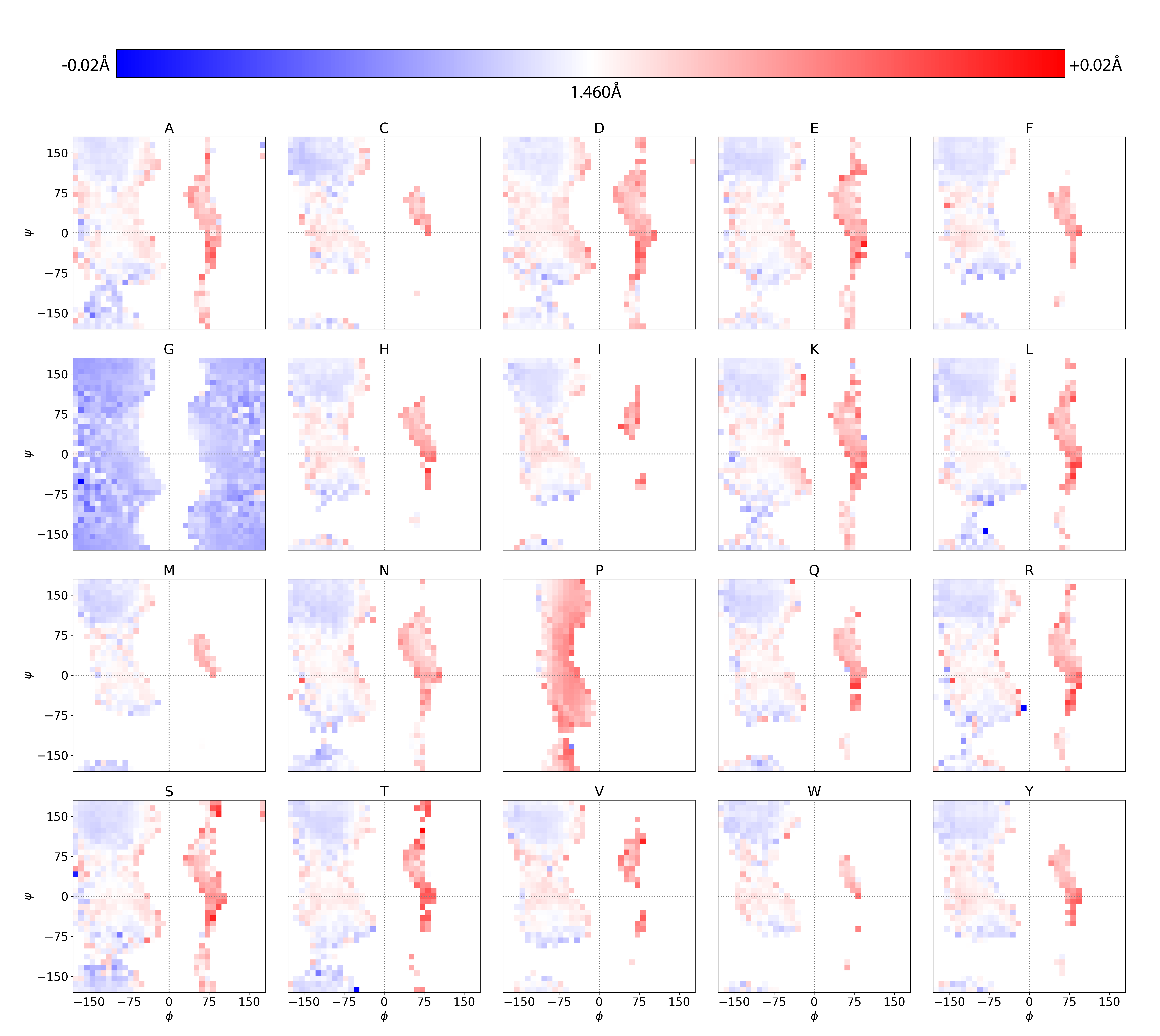}
\end{center}
\vspace{-8mm}
\caption{\textit{$N-C_{\alpha}$ bond length deviations from the mean values averaged over $\phi$ and $\psi$ angles as a function of residue type}. The regions of red correspond to longer bonds while the region in blue show reduced bond values relative to the mean. The $N-C_{\alpha}$ bond lengths were categorized according to $\phi$ and $\psi$ angles rounded to the closest tens.}
\label{fig:d1-res}
\end{figure}

\begin{figure}
\begin{center}
\includegraphics[width=0.98\textwidth]{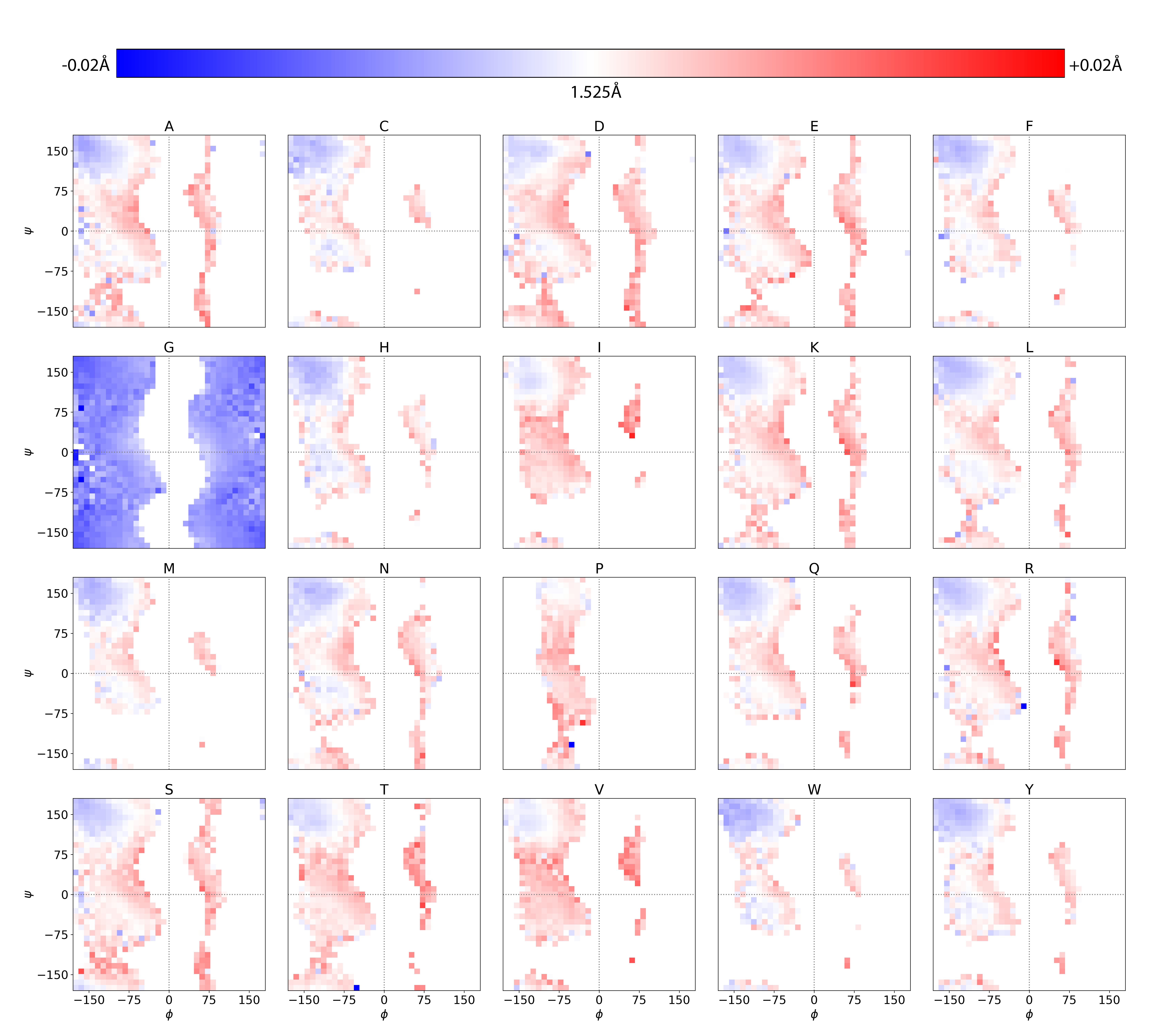}
\end{center}
\caption{\textit{$C_{\alpha}-C$ bond length deviations from the mean values averaged over $\phi$ and $\psi$ angles as a function of residue type}. The regions of red correspond to longer bonds while the region in blue show reduced bond values relative to the mean. The $C_{\alpha}-C$ bond lengths were categorized according to $\phi$ and $\psi$ angles rounded to the closest tens.}
\label{fig:d2-res}
\end{figure}

\begin{figure}
\begin{center}
\includegraphics[width=0.98\textwidth]{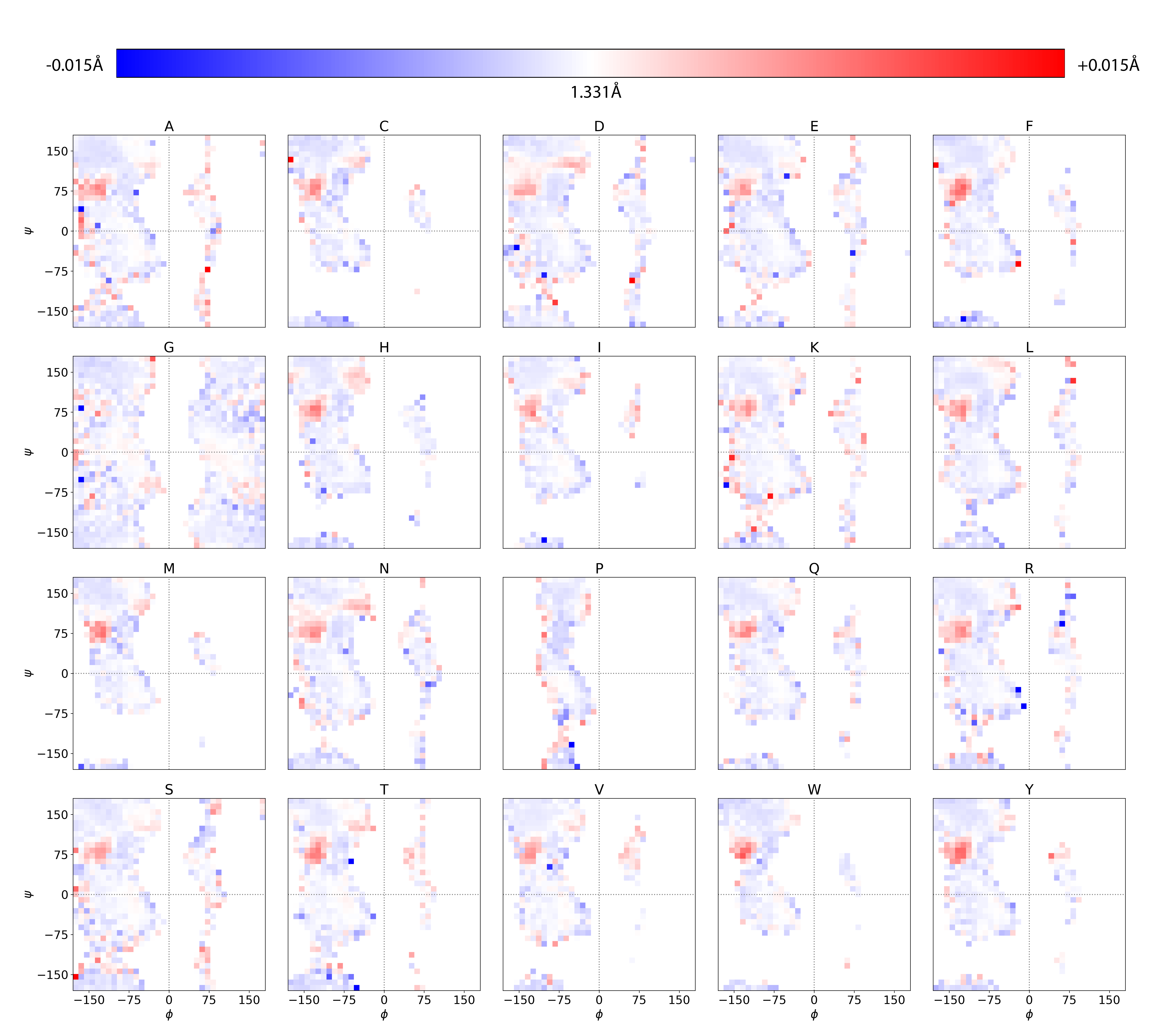}
\end{center}
\caption{\textit{$C-N$ bond length deviations from the mean values averaged over $\phi$ and $\psi$ angles as a function of residue type}. The regions of red correspond to longer bonds while the region in blue show reduced bond values relative to the mean. The $C-N$ bond lengths were categorized according to $\phi$ and $\psi$ angles rounded to the closest tens.}
\label{fig:d3-res}
\end{figure}

\begin{figure}[H]
\begin{center}
\includegraphics[width=0.98\textwidth]{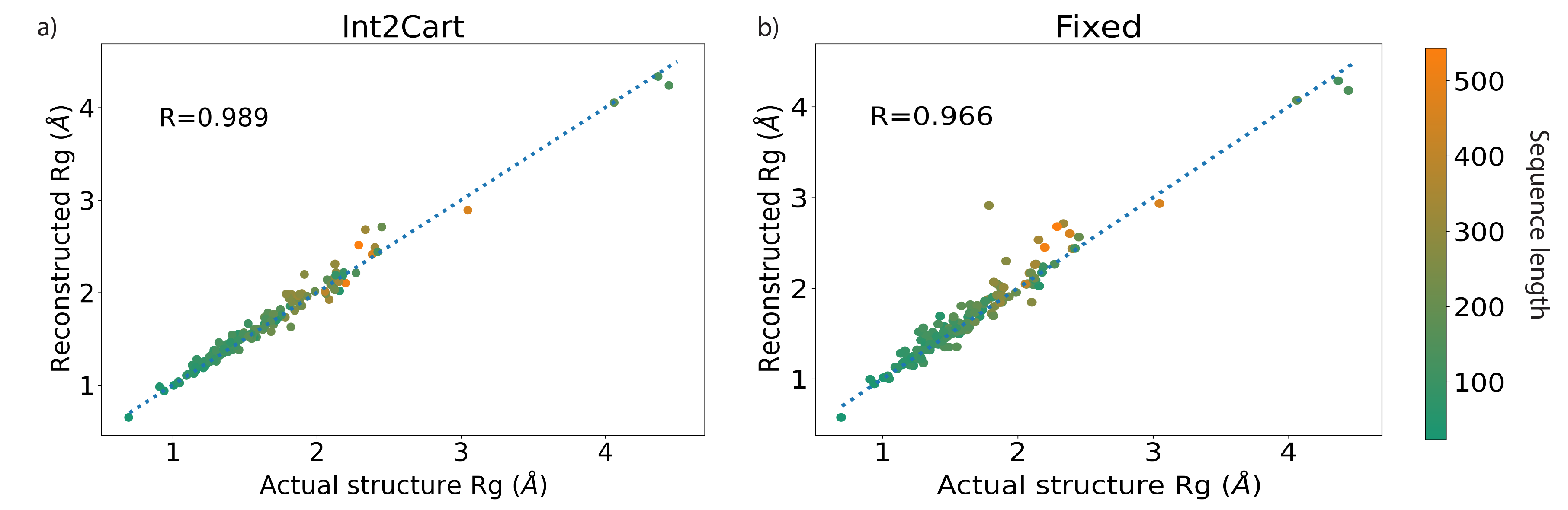}
\end{center}
\vspace{-4mm}
\caption{\textit{The accuracy of radius of gyration when internal coordinates are back-transformed to Cartesian coordinates using Int2Cart or Fixed.} The R$_g$-match calculates the correlation of radius-of-gyration of individual proteins with the reference proteins. Correlations of R$_g$-match values for (a) Int2Cart and (b) Fixed over the test set.}
\label{fig:rg}
\end{figure}

\begin{figure}
\begin{center}
\includegraphics[width=0.98\textwidth]{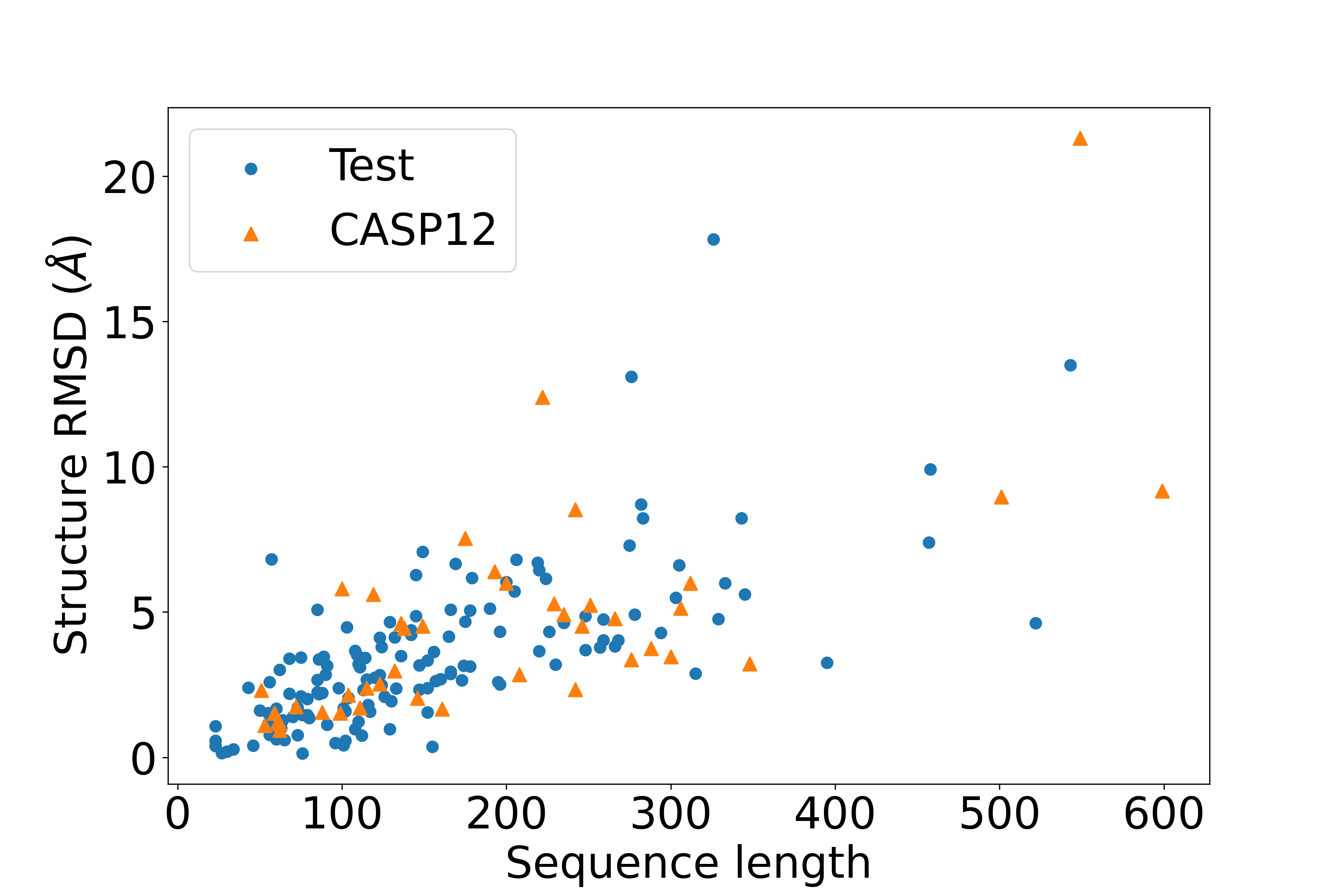}
\end{center}
\caption{\textit{Comparison for reconstructed structure RMSD as a function of sequence length for primary test datset and CASP12 test dataset}. The higher reconstructed structure RMSD for structures in CASP12 test dataset is most likely due to higher proportion of longer proteins compared with results of the primary test dataset.}
\label{fig:d3-res}
\end{figure}
